\PassOptionsToPackage{table}{xcolor}
\documentclass[sigconf, 10pt, superscriptaddress]{acmart}
\usepackage{amsmath}
\usepackage[utf8]{inputenc}
\usepackage[ruled,vlined]{algorithm2e}
\usepackage{hyperref}
\usepackage{syntax}
\usepackage{appendix}
\usepackage{graphicx}
\usepackage[table]{xcolor}
\usepackage{tabularx}
\usepackage{float}
\usepackage{mathtools}
\usepackage{float}
\usepackage{xspace}
\usepackage{collcell}
\usepackage{hhline}
\usepackage{pgf}
\usepackage{multirow}
\usepackage[inline]{enumitem}
\usepackage{makecell}
\usepackage{amsmath}
\usepackage{tikz}
\usetikzlibrary{arrows}

\newcommand*\circled[1]{\tikz[baseline=(char.base)]{
\node[shape=circle,draw,inner sep=2pt] (char) {#1};}}

\renewcommand\footnotetextcopyrightpermission[1]{} 
 \setcopyright{none}

\newcommand{\privateoverall}[0]{2.07\%}
\newcommand{\flashbotoverall}[0]{1.98\%}
 
\newcommand{\totalduration}[0]{12 days} 
\newcommand{\minerpercentageprofit}[0]{65.9\%}
\newcommand{\minerpercentageincome}[0]{9.2\%}
\newcommand{\minerpercentageincomeDeFi}[0]{22.7\%}
\newcommand{\percentagemevprivate}[0]{87.6\%}
\newcommand{\percentageprivateflashbots}[0]{46.5\%}
\newcommand{\percentagemevprivaterounded}[0]{85\%}
\newcommand{\percentageprivateflashbotsrounded}[0]{50\%}

\newcommand{\swapeq}{\underset{\mathrm{swap}}{\equiv}}

\setlist[itemize]{leftmargin=*} \setlist[enumerate]{leftmargin=*}

\begin{document}

\title{Extracting Godl [sic] from the Salt Mines:\\ Ethereum Miners Extracting Value}

\author{Julien Piet}
\affiliation{UC Berkeley}
\email{piet@berkeley.edu}

\author{Jaiden Fairoze}
\affiliation{UC Berkeley}
\email{fairoze@berkeley.edu}

\author{Nicholas Weaver}
\affiliation{UC Berkeley \& ICSI}
\email{nweaver@icsi.berkeley.edu}

 \date{}

\begin{abstract}

Cryptocurrency miners have great latitude in deciding which transactions they accept, including their own, and the order in which they accept them.
Ethereum miners in particular use this flexibility to collect MEV---Miner Extractable Value---by structuring transactions to extract additional revenue. 
Ethereum also contains numerous bots that attempt to obtain MEV based on public-but-not-yet-confirmed transactions. 
Private relays shelter operations from these selfsame bots by directly submitting transactions to mining pools.

In this work, we develop an algorithm to detect MEV exploitation present in previously mined blocks. We use our implementation of the detector to analyze MEV usage and profit redistribution, finding that miners make the lion's share of the
profits, rather than independent users of the private relays. More specifically,
\begin{enumerate*}[label=(\roman*)]  \item 73\% of private transactions hide trading activity or re-distribute miner rewards, and \percentagemevprivate~of MEV collection is accomplished with privately submitted transactions, 
\item our algorithm finds more than \$6M
worth of MEV profit in a period of \totalduration, two thirds of which
go directly to miners, and  \item MEV represents
\minerpercentageincome~of miners' profit from transaction fees. 
\end{enumerate*}

Furthermore, in those \totalduration, we also identify four blocks that contain enough MEV profits to make
time-bandit forking attacks economically viable for large miners, undermining
the security and stability of Ethereum as a whole.

\end{abstract}

\maketitle

\section{Introduction}

The history of traditional financial systems is rife with insider trading.
Market manipulation techniques that exploit information asymmetry---i.e., leveraging
information that has yet to propagate to the public---have adapted to new technologies
enabling digital asset trade. 

The advent of cryptocurrencies like Ethereum offered a permissionless financial platform
in which all users were on a level playing field. 
However, the reality of cryptocurrency-based value exchange is that, akin to traditional
financial platforms, information asymmetry remains. 
As early as 2014, Ethereum miners publicly discussed the possibility of frontrunning,
a behavior that would be illegal in any regulated exchange. 

The concept of extracting value from such information in Ethereum became known as ``Miner
Extractable Value'' or MEV, and the seminal ``Flash~Boys~2.0'' paper \cite{daian2020flash}
demonstrated that MEV extraction was not only possible but widespread. 

Unlike traditional financial systems, legislation for cryptocurrencies is not in place to
enforce good behavior; cryptocurrencies and the larger ``Decentralized Finance'' (DeFi)
ecosystem are largely unregulated platforms.
As a result, adversarial behavior proliferates. 
This poses risks to individual, honest users and has the potential to destabilize the
underlying blockchain.

MEV extraction requires analysis of both the public memory pool and the decentralized exchange platforms. 
Through techniques such as \emph{frontrunning}, \emph{backrunning}, and \emph{sandwiching}, users can create new transactions based on pending-but-uncommitted transactions for considerable profit. 
Since Ethereum blocks are mined about every 13 seconds \cite{ethstats}, adversaries have time to analyze the current memory pool state and potentially identify profitable sequences of new transactions.

More generally, it is not uncommon to see software bugs in Ethereum smart contracts \cite{wohrer2018smart}. 
In the event that funds are stranded in a defective contract, new transactions are often used to ``rescue'' the trapped funds.
Unfortunately, the existence of generalized frontrunning bots \cite{DarkForest1, qin2021quantifying} renders rescue operations impossible without access to private miners.
If a transaction submitted to rescue trapped funds is publicly visible, a bot can see this transaction, copy it, change the address that receives the profits, and offer to pay a higher ``gas price'' (a per-operation transaction fee) to the miners.
As miners naturally select the transactions that pay them more, they will preferentially select the bot's transaction, allowing the bot to claim the funds for itself.  
MEV-conscious users may opt to preemptively raise the gas price of their own transactions, effectively paying an additional fee on top of transaction fees to get stronger finality guarantees. 
In essence, MEV extraction drives up the gas price paid by regular Ethereum users.

To avoid detection by bots, users can opt to send transaction data directly to miners, bypassing the public memory pool entirely. 
This method would also be useful for bots seeking to avoid competition with other bots.
Transacting directly requires having a private communication channel with a miner, something most users do not have.
Flashbots Auctions \cite{flashbots} pioneered such a service to enable private MEV extraction, creating a private link between a user or bot and a participating miner. 
The authors claim Flashbots removes the burden of MEV extraction traffic from the public memory pool and mitigates MEV's negative externalities, allowing for fair access to MEV opportunities.

Still, miners are in a favorable position with access to both public and private information that might allow them to detect profitable opportunities before other users. Furthermore, miners constructing or supporting MEV also have an advantage in atomicity.  
For a normal transaction, the transaction itself is atomic, but from a miner's viewpoint, the entire series of accepted transactions in a block is committed in a single atomic action.  
This property undermines the notion that Ethereum and the DeFi systems built on top of Ethereum provide a fair and permissionless system placing all users on a level playing field. 



In this work, we employ an empirical approach to study \emph{past} MEV extraction: the value extractable by manipulating \emph{transaction order}. We investigate the use of private transactions and analyze the effect of Flashbots on MEV. 
We find that today, most MEV extractions use private transactions. 
However, despite Flashbots and other MEV relays' objective of fair access and risk mitigation, MEV profits are largely claimed by miners, and existential stability risks to the blockchain are more present than ever.

\paragraph{Roadmap}
First, we give an overview of our contributions in Section~\ref{sec:contributions}. 
In Section~\ref{sec:background}, we provide background on Ethereum, DeFi, and the concept of MEV.
Section~\ref{sec:related} gives an overview of related works in the field.
We then discuss private transactions and MEV relays in Section~\ref{sec:private} before describing our methodology and MEV detection algorithm in Section~\ref{sec:experimental}.  
We present our experimental findings in Section~\ref{sec:results}.
Finally, we provide insight into possible countermeasures and the impact of Ethereum's shift to proof-of-stake in Section~\ref{sec:discussion}.

\section{Contributions} \label{sec:contributions}
In order to study historical MEV extraction through private transactions, we implement a
custom Ethereum node that captures transaction-related data in real-time and checks every
transaction against the memory pool before tagging it as private. 
Previously, MEV detection used point-wise heuristics on historical data or graph methods
to find arbitrages in decentralized exchanges (DEXs).  
We develop a generic arbitrage detection technique that, to the best of our knowledge, is
the first method to identify generic MEV opportunities in historical data.
By finding specific cycles in cryptocurrency transfer graphs across blocks, our algorithm
can detect arbitrage, backrunning, and frontrunning, even across multiple transactions. 
We summarize our contributions as follows: 
\begin{description}
    \item[Private transaction analysis.]  
    Private transaction usage is inconsistent across miners.
    Most private transactions are used for miner profit redistribution, and more
    importantly, MEV exploitation. 
    In fact, in our data, over \percentagemevprivaterounded~of MEV is extracted using
    private transactions. Flashbots accounts for under
    \percentageprivateflashbotsrounded~of private transactions---the rest
    are either owned by the miner or submitted through covert channels.  
    
    \item[Miner profit analysis.]  
    Despite Flashbots' claim to a fair MEV opportunity redistribution, miners dominate the
    market: they make up two thirds of all MEV-related profits, and most likely control
    many bots themselves. In our data, this represents on average \minerpercentageincome~
    of their total transaction fee income and
    \minerpercentageincomeDeFi~of their income from decentralized finance
    applications. 
    As for private transactions, MEV extraction is inconsistent among miners---the top
    five miners earn more from MEV than all bots combined, but 40\% of miners do not
    partake in MEV extraction.  
    
    \item[Concrete time-bandit attack opportunities.]  
    Previous work has warned of the risk for time-bandit attacks if MEV profits were to
    rise, where major miners are incentivized not to add new transactions but instead,
    attempt to rewrite old history to capture the present MEV for themselves.
    Some even found hypothetical opportunities that would have yielded enough profit to
    warrant an attack.  
    Using game-theoretic models developed in \cite{gervais2016security}, we found four
    blocks in \totalduration~that earned enough to warrant an attack from miners with a
    hash rate superior to 10\%.
\end{description}

\section{Background}
\label{sec:background} 

Ethereum \cite{vitalik2014ethereum} is a decentralized blockchain that allows
users to create ``smart contracts.'' 
Abstracting away security concerns, a blockchain is a shared global ledger that is implemented using an append-only data structure. 
Sets of transactions are grouped into blocks and added to the ledger as a hash chain, with each block containing the hash of the previous block to create a linear, append-only, tamper-resistant data structure.

The blocks themselves are decided by miners---special nodes that choose a set of pending transactions from a larger pool, check their consistency, and submit the block to the network along with a proof of validity. 
Once other nodes accept that a block is valid, they move on to creating the next block from the remaining unverified transactions.

For Ethereum, the proof of block validity is a ``proof of work'': the miners attempt to generate partial hash collisions, and a block with a significantly low hash value is then broadcasted to all other mining nodes and accepted as the new head of the blockchain.

\begin{figure}[ht] 
\centering
\includegraphics[width=\linewidth]{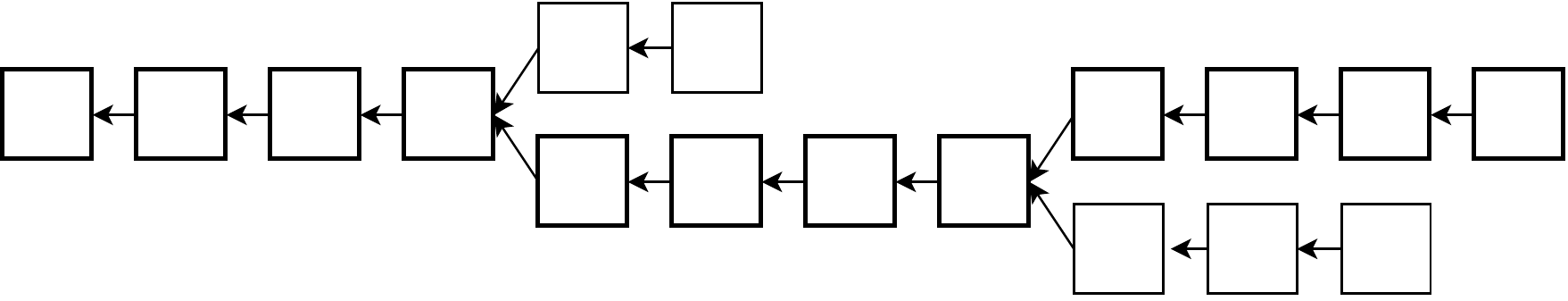} 
\caption{
When multiple nodes are created at the same time, it is possible for the chain to fork. In this case, the valid chain is the longest one.}
\label{fig:forking} 
\end{figure}

It is, however, possible for multiple blocks to be created at the same time and point back to the same parent block, either because two miners solved the problem at the same time, or for adversarial reasons. 
The convention is that honest nodes should always choose the longest chain as valid. Figure~\ref{fig:forking} provides an illustration of such a scenario. 

Miners participate in this scheme because each new block includes a block reward: a payment in cryptocurrency to the miner which produces a valid proof. 
In Ethereum, this reward consists of a static block reward, inflating the total supply of Ethereum and fees associated with each accepted transaction.

Older cryptocurrencies like Bitcoin \cite{nakamoto2008bitcoin} only process basic transactions, but Ethereum transactions are coupled to a small stack machine called the Ethereum Virtual Machine (EVM), and transactions represent small deterministic programs.  
Each EVM instruction in a transaction costs a fixed amount of ``gas,'' and the transactions themselves include a gas fee: an amount of Ethereum they will pay for each unit of gas consumed.  

Ethereum uses this virtual machine to support ``(smart) contracts.'' 
These are stored internally as a user; node memory additionally stores executable bytecode on top of maintaining balances for this user. 
One can think of a smart contract as a continuously available program---every interaction with the program will run a function, which can update balances, modify the program memory, or even call other contracts. 
Users interact with contracts by sending transactions with input code (as well as cryptocurrency) interpreted by the contract to run a function.

\begin{figure}[ht] 
\centering
\includegraphics[width=\linewidth]{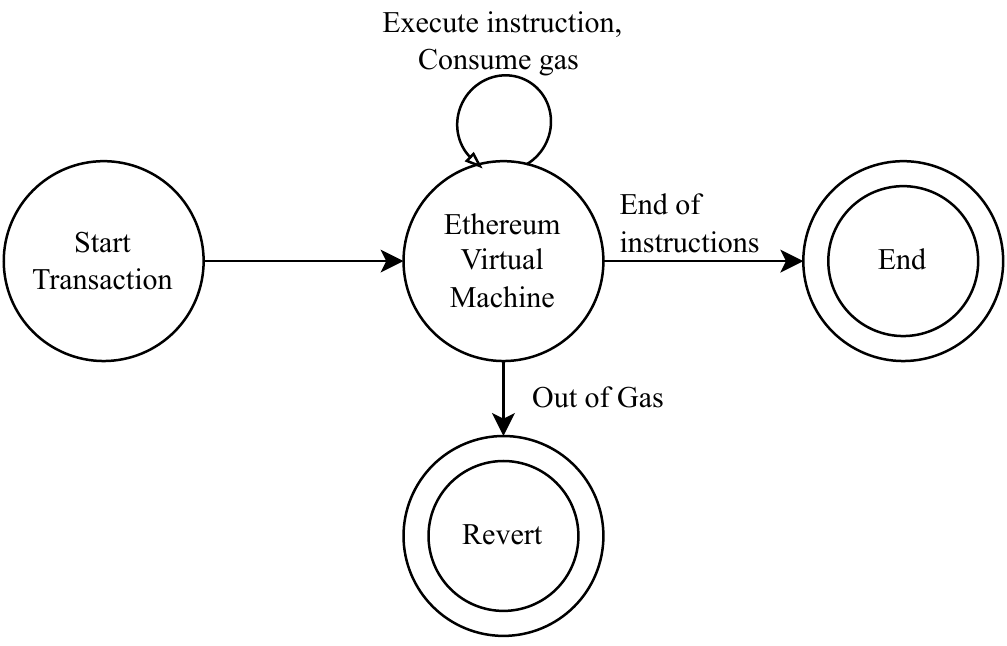} 
\caption{
Smart contract execution.}
\label{fig:evm} 
\end{figure}

The deterministic nature of smart contracts ensures that all miners should compute the same value given the same state, and the related gas limit allows the system to avoid the halting problem: a program can ``run out of gas'' with the fees collected by the miner while not actually performing any useful work. 
The process of executing a contract is shown in Figure~\ref{fig:evm}.

The ability to create such contracts has lead to an entire financial ecosystem running on
top of the EVM.  
Standard contracts can create additional cryptocurrencies as distinct ``tokens'' with the contract's internal memory keeping track of user balances.  
Other smart contracts implement exchanges, maintaining pools of tokens where users can buy and sell.
Since these programs themselves are running on demand on every Ethereum validator, the developers have dubbed the larger system as ``DeFi'' or ``Decentralized Finance.''

We now provide useful details on crucial parts of the Ethereum ecosystem.

\paragraph{Transactions on Ethereum} 
By default, Ethereum users broadcast transactions via gossip through a peer-to-peer network. 
They specify a gas price that indicates how much they are willing to pay for a unit of computation on the EVM. 
Miners receive pending transactions, execute them, and forward the resulting state change to the network. 
To do this, miners generally order transactions by descending gas price. 
(In practice, the sorting approach can vary across different Ethereum clients.)

There are multiple types of transactions: 
\begin{enumerate*}[label=(\roman*)]
    \item \textit{regular} transactions from one address to another,  
    \item \textit{contract deployment} transactions where additional data is used as the new contract bytecode, and  
    \item \textit{contract execution} transactions, where the recipient is a contract that executes input data from the transaction.
\end{enumerate*}

\paragraph{Private transactions} 
Ethereum users may opt to bypass the public transaction pool entirely and directly liaise with miners to get transactions included in a block.  
Transactions confirmed in this manner are known as \textit{private transactions}.

\paragraph{Transaction fees} 
Since the EIP-1559 Ethereum Improvement Proposal \cite{eip1559}, transactions burn part of the gas (that is, remove ETH from the system to create deflation).
Each block has a ``base fee per gas'' which quantifies how much ETH will be burnt per unit of computation. 
Transaction must at least pay this gas price. 
However, by paying more, the remainder of the gas is earned by the miner as a gas "tip."
This tip is what we consider to be the miner's remuneration from transactions. 

\subsection{Decentralized Exchanges (DEXs)}

A decentralized exchange (DEX) is a distributed marketplace that allows users to swap tokens for different tokens. 
Critically, DEXs are untrusted in the sense that users do not need to trust the DEX itself for security---it is only trusted for availability.
DEXs achieve this reduced trust by leveraging decentralized blockchains such as Ethereum.
DEXs are usually constructed from smart contracts; the largest DEX at the time of writing, Uniswap \cite{adams2021uniswap}, is a prime example.
Order book matching---the process of matching buy and sell orders---can be encoded into smart contracts that run directly on the decentralized blockchain.
Since the smart contracts execute on a decentralized blockchain, a number of desirable properties emerge. 
First, the contract itself is publicly visible, and anyone is free to inspect its correctness. 
Second, The execution of the contract cannot deviate from its program specification so long as the underlying blockchain (e.g. Ethereum) is secure. 
In other words, it would take a miner with 51\% hash rate (to carry out a ``51\% attack'') to arbitrarily control smart contract execution---it is believed that such an attack is prohibitively expensive on Ethereum.
In contrast, centralized exchanges that must be trusted to carry out order matching in a fair manner since users have no way to confirm honest behavior.

DEXs pose new design challenges relative to centralized exchanges. 
Since pending transactions can be inspected on the public blockchain, MEV extraction is commonplace. 
Similarly, large transactions can impact prices substantially, causing \emph{slippage}: significant price changes between when a transaction is submitted and finalized. 
These are not as critical in centralized exchanges since the central entity privately handles order flow, thus making it more difficult for bots to predict slippage.\footnote{In conventional finance, the exchanges in the past would take advantage of this but such practices were banned following the Great Depression.}

We summarize the high-level trade-offs between DEXs and centralized exchanges as
follows: 

\begin{description} 
    \item[Privacy.] DEXs do not enforce a ``know your
    customer'' policy that binds an exchange to gather personal details about each customer to prevent money laundering and other criminal activities.
    
    \item[Security.] Security of DEXs usually reduces to the security of the underlying blockchain. Security can be unverifiable in many centralized exchanges since they are almost always closed services.
    
    \item[Trust.] In DEXs, swaps are handled by AMM protocols (discussed below) which are decentralized in trust, i.e., users need not trust a central entity to democratically match orders.
    
    \item[Regulation.] Centralized exchanges are covered by legislation that ensures some degree of honest behavior (both by the exchange itself and its users).
\end{description}

\paragraph{Automated Market Makers (AMMs)}

Automated market maker (AMM) refers to the underlying protocol that supports a DEX. 
They can be thought of as the algorithmic procedure that allows DEX users to swap tokens without involving a third party. 
In contrast to traditional market makers for centralized financial platforms, AMMs must achieve similar functionality without relying on any trusted intermediaries. 
To ensure a seamless trading process, AMMs must ensure that liquidity is high---it should be \emph{easy} to buy or sell a token. 
Liquidity is inversely correlated to slippage: if liquidity is low, the price of an asset might change significantly in the time it takes to initiate and finalize a trade and vice versa. 
For example, a large buy order could quickly increase the value of a token.

AMMs typically use a liquidity pool to provide liquidity for specific token pair. 
That is, the AMM itself holds both tokens that it offers in a trading pair and, if a vulnerability is discovered, these tokens are often stolen by attackers.  

The tokens involved in a trading pair are sourced from users that deposit funds into the pool.
Such deposits should include both funds at some pre-determined ratio. 
AMMs ensure a balanced ratio of assets in liquidity pools by encoding balancing equations into the underlying smart contract. 
Uniswap, for example, uses a linear equation to set the mathematical relationship between trading pairs in liquidity pools \cite{adams2021uniswap}. 
This system is particularly susceptible to slippage---DEXs rely on independent arbitrageurs who quickly fill supply discrepancies across exchanges, thereby providing liquidity. 

\subsection{Miner Extractable Value (MEV)} 
Miner extractable value (MEV) originally denoted the value miners can leverage from smart contracts beyond the standard block reward and gas fees. 
This usually manifests as the value gained by modifying the set or order of transactions in a block.
However, this term has evolved to mean \emph{maximal} extractable value, which encompasses all value that can be extracted from the blockchain, both by miners and other parties.

In the proof-of-work context, miners are responsible for the ordering of transactions within a block, and thus have the power to implement MEV strategies.
The miners also have an advantage in atomicity. 
For users, a transaction is atomic: either an exchange happens or it doesn't.  
However, for the miner, the unit of atomicity is the block: either all transactions happen or none do.

However, users can also manipulate the placement of transactions in a block by changing the gas fee, or they can collaborate with miners using private communication channels to profit from MEV opportunities.

Most prior work only considers the case for bot-driven MEV \emph{without} collusion with miners.  
In this work, we additionally analyze miner collusion through private relays, and miner-orchestrated MEV.
Previous work has identified four main forms in which MEV can manifest:

\begin{description}
    \item[Frontrunning] is a concept from traditional finance. 
    It consists of capitalizing on knowledge of a pending transaction before it has been included in a block.  
    Possible strategies can be to suppress the transaction, duplicate it, or submit your own order in advance.
    The most common forms of frontrunning on Ethereum are \emph{sandwich} attacks in which the adversary uses the price slippage caused by a large trade to buy some assets right before the transaction and sell them right after.  
    In our work, we define sandwich attacks as any MEV trading strategy that exchanges
    a set of tokens over two separate transactions in a block.

    \item[Backrunning] is similar to frontrunning---the knowledge of a
    transaction is used to insert an order right \emph{after} the target transaction.  
    In most cases, the price slippage due to a transaction will change the exchange rate on one exchange, but not on others.
    Using this information, backrunners exploit the price difference between multiple DEX exchanges to turn a profit.  
    In this work, backrunning extracts MEV through a single transaction in which the same
    cryptocurrency is exchanged on multiple DEXs simultaneously, perhaps in different quantities. 

    \item[Arbitrage] is a strategy in which an adversary buys and sells a set of
    tokens in different markets for a profit.
    It leverages market inefficiencies by exploiting subtle differences in exchange rates.  
    In this sense, both backrunning and sandwiching are forms of arbitrage in which the market inefficiency is generated by a large transaction.  
    Often, users can leverage flash loans to gain more out of an arbitrage opportunity.
    Flash loans are large loans only valid for a single transaction---they are
    contingent on the user being able to repay them at the end of the
    transaction with an added fee.  
    If the loan cannot be repaid, the transaction is voided, and the lender gets their assets back.
    
\end{description}

\section{Related Work} \label{sec:related}

\cite{daian2020flash} first introduced the concept of MEV.
This paper was the first, to our knowledge, to analyze bot competitions to exploit MEV opportunities. 
They uncovered ``Priority Gas Auctions'' (PGAs) where bots resubmit pending MEV transactions at higher gas fees to beat their competitors.  
PGAs threaten blockchain usability: they are responsible for heavy network congestion and increase gas prices. 
However, the main threat on which this paper focused is directly linked to MEV.
They claimed high-profit MEV opportunities might entice miners to fork the chain, threatening Ethereum's stability.
This work spawned a new line of research into MEV strategies, both for miners
and bots, as well as potential countermeasures against MEV extraction.
Most notably, the Flashbots MEV relay \cite{flashbots} was created in order to mitigate some of these risks by removing public PGAs in the hopes of reducing network congestion and making MEV easier and fairer to extract.

MEV itself has been discussed in many works since Flash Boys 2.0.  
The broadness of the activity captured by MEV varies depending on the work, ranging from the value extractable \emph{for the miner} by reordering transactions \cite{Judmayer2021Estimating} to the total value that can be extracted by reordering, adding, or deleting transactions regardless of the party \cite{zhou2021just}.
Recent studies have even extended the definition to multi-chain MEV, where value can be extracted from transactions involving multiple blockchains \cite{obadia2021unity}.

\paragraph{MEV Search and Analysis}

A line of research focuses on identifying specific forms of MEV, both online
(looking at current DEX graphs and pending transactions) and offline
(looking at historical activity). 
\cite{zhou2021just} leverages DEX pricing information to find arbitrage loops using negative cycle detection in currency exchange graphs. 
Their strategies would have generated 191.48 ETH per week on average, if executed. 
\cite{wang2021cyclic} analyzes past cyclic arbitrage, and shows that potential MEV profits from the popular DEX Uniswap are rapidly increasing.

\cite{torres2021frontrunner} provides a methodology for identifying three specific types of frontrunning attacks on the Ethereum memory pool: displacement, insertion, and suppression. 
They measure their historical impact over five years. 
Of the three aforementioned types, only insertion has the potential of generating more MEV since it leverages the arbitrage opportunity created by price slippage due to a third party transaction.
In our categorization, both sandwiching and backrunning attacks correspond to insertion tactics.  
Displacement captures duplicating another user's profitable transaction and executing it before they do. 
This does not change the total amount of MEV generated, but instead, changes the entity who profits from it. 
Suppression aims at barring another transaction from being mined in a block, which again does not generate any extractable MEV itself.  
\cite{eskandari2019sok} provides a complete overview of frontrunning attacks which are not limited to MEV extractions.

\cite{zhou2021high} focuses on building sandwich attacks against memory pool transactions and shows that even with competing adversaries, the attack remains profitable.  
\cite{zust2021analyzing} measures past opportunities for sandwiching attacks and shows that the number is increasing.

Finally, \cite{qin2021quantifying} provides an economic analysis of MEV (or as they coin it, \emph{Blockchain Extractable Value}) in past blocks.  
They use separate heuristics to find sandwiching, arbitrage and liquidation profits, as well as propose a new automated technique for running transaction duplication attacks. 
This technique is similar to the displacement attack in \cite{torres2021frontrunner}.  
They also provide an analysis of MEV relays such as Flashbots and show that, contrary to its intentions, it does not significantly reduce network congestion between Ethereum nodes---our work is closest to this effort.  
While we conduct an analysis on a shorter time span, our solution provides a unique detection algorithm for a broader class of MEV than captured by \cite{qin2021quantifying}. 
We further provide insight into the provenance of such MEV exploits, tracking profit redistribution to miners.

\paragraph{Private Transactions}

Another consequence of the original paper on MEV \cite{daian2020flash} is the increased usage of private transactions through MEV relays such as Flashbots.
As mentioned above, \cite{qin2021quantifying} analyzes MEV relays and finds that they do not reduce network congestion. 
\cite{capponi2022evolution} provides empirical evidence that MEV relays (more broadly coined \emph{dark venues} in this work) do not mitigate frontrunning risks or reduce transaction costs.
They find that MEV relays increase miner payoff---a conclusion we support in this work by analyzing miner profits from MEV.

The concept of dark pools extends beyond cryptocurrencies: private transactions
also exist in traditional finance.  
\cite{zhu2014dark} builds a theoretical model for financial dark pools. 
They show that they impact the overall fairness, information symmetry and transparency of the exchange market.
These findings further apply to Ethereum private transactions.

\paragraph{Miner's perspective}

MEV is called miner extractable value for a good reason: miners are those that
can directly exploit it in proof-of-work blockchains.  
In our work, we show that miners make the vast majority of profits, even with MEV relays.  
Previous works have focused on strategies for miners to maximize this revenue.
\cite{angeris2021note} formalizes optimal bundle ordering of transactions by modeling them as an integer linear program. 
\cite{gervais2016security} discusses the incentives for forking attacks for profit maximization.  
The authors use Markov Decision Processes to model optimal adversarial behavior. 
We leverage their analyzes to quantify forking risks related to MEV in our data.

\paragraph{Countermeasures}

MEV is a threat to the stability and performance of the blockchain. 
This is emphasized by many papers on the topic \cite{daian2020flash, zhou2021just,qin2021quantifying, wang2021cyclic}.  
It creates network congestion, increases transaction prices, increases the cost of participation in DEXs, and most importantly, threatens blockchain consensus.  
A new line of work focuses on potential countermeasures to mitigate these risks.

One strategy is to change the model used to create a DEX. 
\cite{mcmenamin2022fairtradex} proposes FairTraDEX, a new platform based on frequent batch auctions instead of AMMs. 
They provide formal game-theoretic guarantees against MEV extraction. 
\cite{heimbach2022eliminating} proposes a new algorithm for setting the slippage tolerance of swaps to prevent sandwich attacks.
\cite{ciampi2021fairmm} introduces off-chain communications to ensure fair trades.  \cite{baum2021p2dex} used multi-party computation to create a privacy-preserving cross-ledger exchange.  
\cite{zhou2021a2mm} advocates
for a unified AMM for the blockchain to mitigate sandwich and backrunning attacks.

Another line of research focuses on order fairness. 
By having consensus on the ordering of a transaction, neither individual miners nor bots can modify the block to extract MEV.  
\cite{byers2022combating} gives an overview of ordering consensus. 
\cite{kelkar2020order} proposes a new class of protocols, Aequitas, which achieves order-fairness in a decentralized manner.
However, they only enforce a weak version of order-fairness, coined \emph{batch order-fairness}, which does not prevent all MEV. 
\cite{kelkar2021themis} provides similar results to Aequitas with their Themis scheme. 
\cite{kursawe2020wendy} proposes Wendy, a set of protocols for insuring order-fairness based on ordering linearizability.
Unfortunately, this only provides a coarse order-fairness result, and does not
mitigate all reordering possibilities.

Finally, some papers suggest encrypting transactions and only revealing their content after a consensus is reached on the list and order of transactions included in the next block.  
\cite{yakira2021helix} proposes a novel distributed ledger consensus protocol which uses threshold encryption to hide the content of transactions before a consensus is reached and enough honest parties agree to proceed.  
Their solution is elegant and provides strong privacy guarantees; however, (1) it requires a more complex validation procedure than Ethereum, (2) needs an impractical number of network nodes, and (3) fails to prevent the elected leader from adding some of its own encrypted transactions to the start of the block.

\section{Private Transactions} \label{sec:private}

In the vast majority of cases, users who wish to submit a transaction to the Ethereum blockchain must send the transaction to a mining node. 
This is done using the Ethereum Wire Protocol which codifies how Ethereum nodes communicate and share state. 
Users start by submitting their transaction to a node, which in turn broadcasts the  transaction to other nodes until the transaction gets mined (and exits the pending transaction pool) or expires. 
However, some transactions bypass the memory pool and are directly included in blocks: this can only happen if the miner includes its own transactions or if the transaction emitter contacted the miner through private channels.

Private transactions are typically useful in the following cases:
\begin{description} 
    \item[Miner payments.]  
    By including Ethereum transfers in their own blocks, mining pools can avoid paying a gas tip to another miner when re-distributing block rewards to pool members.  
    Without a gas tip, these transactions would not be mined if submitted to the memory pool.
    
    \item[Hacks.]
    When exploiting a smart contract flaw, hackers have a better chance of success if they collude with a miner to bypass the memory pool.  
    If they submit their hack to the memory pool, they run the risk of being detected before being included in a block and having another user steal the opportunity.  
    
    \item[Rescue operations.] 
    Conversely, when white-hat hackers find a flaw in a smart contract with unprotected funds, it is safer for the contract owner to recuperate these funds with a private transaction.
    Public transactions can raise suspicion of exploit bots that might try to steal the funds.  
    A notable example of this is illustrated in \cite{DarkForest1}, which led the same team to use a private transaction in a future rescue attempt, described in \cite{DarkForest2}.
    
    \item[MEV extraction.] 
    MEV is highly dependent on transaction order: arbitrage opportunities are only valid for a specific set of exchange rates between pairs of coins.
    Every transaction involving one of the coins in each pair will change the exchange rate, which in turn could destroy the arbitrage opportunity.
    Similarly, frontrunning and backrunning are both dependent on placing one's transactions as close as possible to a large exchange order.
    Using private transactions allows the user to bypass the PGA for public transactions and directly ask the miner to place the transaction at a specific index for a fee.
    Furthermore, submitting MEV-extracting transactions to the memory pool increases the risk of having the opportunity stolen by an adversarial bot.
    MEV relays, such as Flashbots, create private communication channels between users and miners in order to bypass the memory pool. 
    We describe these systems in more detail in the next section.
\end{description}

In order to better understand the use cases for private transactions and their effect on the blockchain, we designed methods to identify private transactions in mined blocks. 
One technique consists of looking for gas ordering inconsistencies; however, some miners fake the gas tip on private transactions in order to masquerade them.
Instead, we modified a go-ethereum \cite{geth} node to dump all advertised transactions over the network to a file which we then used as a checkup list for transactions in blocks.

One potential drawback of this technique is that in the time between blocks---around 13 seconds---we have no guarantee that the transaction broadcast
has converged.
That is, we may receive pending transactions after they were included in a block, or worse, not receive them at all.
In fact, due to the decentralized nature of the underlying protocol, every node has its own memory pool and, unlike the blockchain itself, its integrity is  \emph{not} ensured by a consensus algorithm---each memory pool might be slightly different and we cannot guarantee that we capture all pending transactions.

\begin{figure*}[ht] 
\centering
\includegraphics[width=\linewidth]{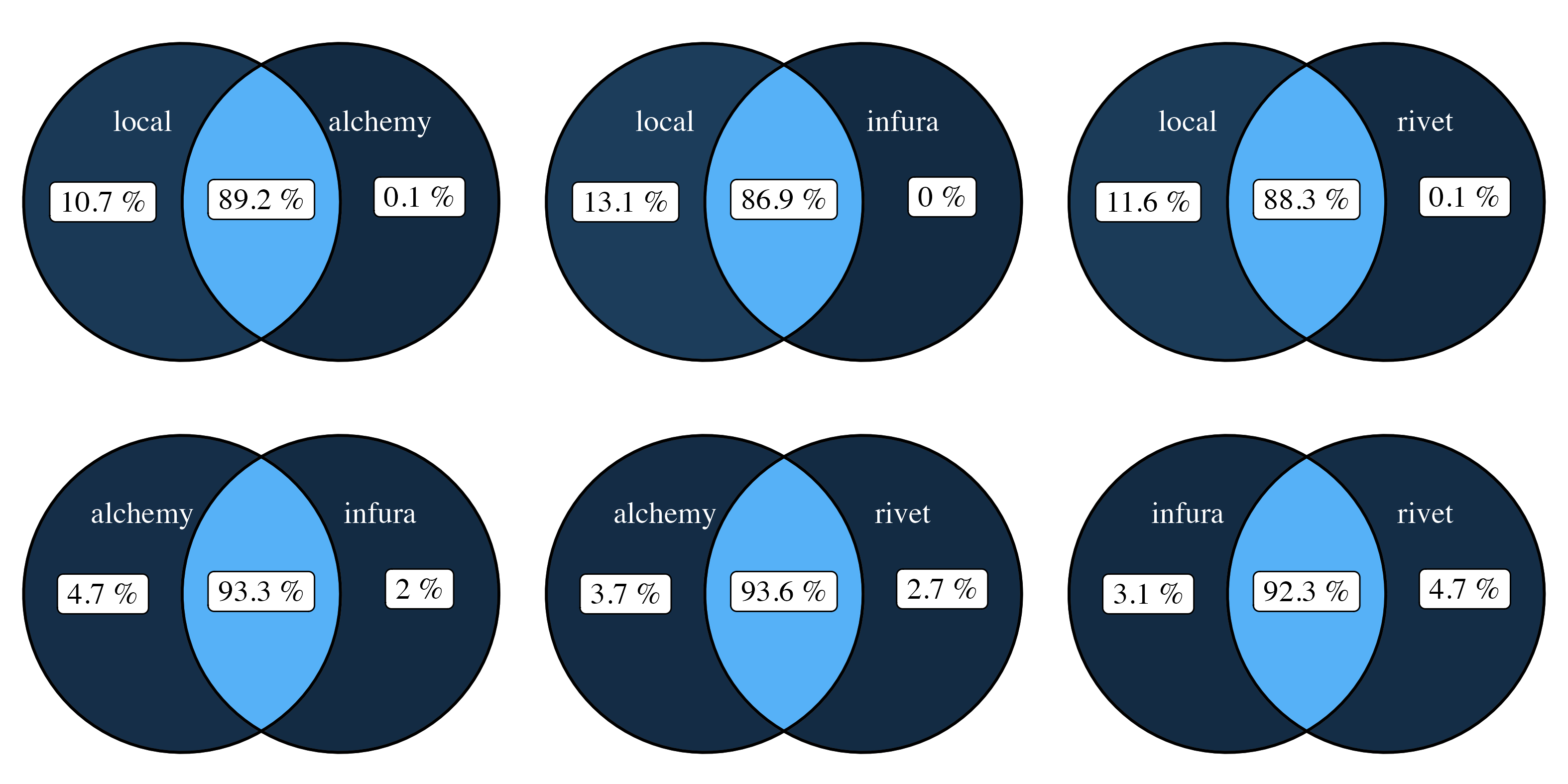} 
\caption{
Intersection percentage between memory pools from three public nodes and our
own node.}
\label{fig:txpool-overlap}
\end{figure*}

In order to measure this effect, we created accounts on three online Ethereum nodes across the world and set up our own measurement device in the United States.
We recorded the transaction pools from all four sources from February 15, 2022 9 AM GMT to February 16, 2022 6 PM.  
We then compared these pools and looked at their intersections.
Figure~\ref{fig:txpool-overlap} gives the size of each intersection.

Our local node's better performance can be attributed to the fact we did not filter low gas transactions.
We also used dedicated hardware instead of an online service. 
Nevertheless, only 0.1\% of transactions from other nodes did not appear locally, giving us confidence that we captured most transactions submitted to the memory pool.
We used our local node in our latter experiments to identify private transactions.

\paragraph{Flashbots Auction}

In a nutshell, Flashbots Auction is designed to replace the existing way Ethereum users communicate transactions to miners. 
In base Ethereum, transactions are frontrunnable because adversaries can analyze the public memory pool. 
Flashbots Auction is designed to mitigate MEV extraction by providing a \textit{private} transaction pool that allows miners to learn optimal block constructions without alerting third parties to MEV opportunities.  
Flashbots Auction uses a first-price, sealed-bid auction which allows users to privately communicate their bid to miners.  
This auction maximizes miner payoff and makes MEV opportunities explicit---this mechanism is intended to reduce MEV exploitation by shedding light on the process.

Flashbots Auction defines three entities: 
\begin{enumerate*}[label=(\roman*)]
    \item \textit{searchers} who intend to use Flashbots' private transaction pool over the standard peer-to-peer transaction pool for transaction submission, 
    \item \textit{relayers} who propagate bundles received from searchers, validate them, and forward them to miners, and 
    \item \textit{miners} who collect bundles and produce a block. 
\end{enumerate*}

The searchers' goal is to bundle transactions in a particular order.
They send this bundle directly to the relayer, thereby bypassing the public memory pool. 
Searchers express their bids via transaction gas price or by directly transferring ETH to the miner's address. 
The relayer, if acting honestly, validates the transaction bundles and forwards them to miners. 
Relayers can provide additional services such as bundle merging and various execution services. 
Such execution services allow searchers to specify the type of transactions they intend to bundle.
Critically, in order to perform their role, relayers must observe transaction details within the bundle.
Thus, a malicious relayer can perform frontrunning.
Once the bundle gets to a miner, they simply attempt to mine a block containing the transactions in the specified order. 
Unlike standard Ethereum clients, Flashbots Auction miners do not necessarily order transactions by gas price, i.e., fine-grained transaction ordering is supported to maximize miner value extraction.

The main issue with Flashbots Auctions is that relayers can inspect transactions
and therefore need to be trusted to not exploit information asymmetry.  

\paragraph{Flashbots Data}

Flashbots publicly releases all mined bundles at the following address: \url{https://blocks.flashbots.net/}. 
This data is accessible in full and contains the list of blocks and transactions involved. 
It also includes any miner fees paid in gas or as separate transactions.  
We downloaded this data and processed it to tag all Flashbots transactions in our analysis.


\paragraph{Other MEV relays}
Following \cite{daian2020flash}, other systems surfaced with the goal of democratizing MEV extraction on Ethereum.
The main competitor to Flashbots is the Eden Network \cite{piatt2021eden}.
This network uses the concept of \emph{slot tenants}: by putting transaction positions within a block ``up for auction,'' Eden Network users can explicitly request for a specific position at a fee.
Like Flashbots, this structure does not remove the possibility of frontrunning.
Miners still have the power to view transaction information and can therefore be bribed to reorder transactions.
Alternatively, miners can conduct MEV extraction themselves.


\section{Experimental setup} \label{sec:experimental}

Our goal was to analyze private transactions and the MEV exploited through them.
In order to do so, we had to save all transactions seen in the memory
pool, but also fetch the content of Ethereum blocks, transactions, and logs to
retrieve DEX currency transfers and exchanges.  In addition, we also needed to
retrieve internal transactions to obtain hidden transfers.

The best way to do this was to set up our own Ethereum node which, as shown
above, could save memory pool transactions with tolerable loss and accomplish
all the other goals.  In particular, we needed an Ethereum node in order to
instrument the virtual machine and get all transfers.  We chose not to use an
archival node since it requires a lot of fast storage---instead our analysis ran live.

We wrote a graph-based algorithm to find MEV extractions in existing
transactions and trace their profits.  Our analysis is based on data collected from February 12, 2022 to February 24, 2022, with a total of 13,898,577 transactions and
74,528 blocks. We have two small data gaps of around an hour on February 15 and
February 17 due to network outages.

We now explain how we modified an Ethereum node to save transactions and track
transfers. Then, we go into more detail about our graph algorithm. 


\subsection{Modified Node}


We installed a slightly modified go-ethereum (also known as geth) \cite{geth}
node on an Ubuntu 20.04 LTS based in the US west coast. We modified the
following:  
\begin{itemize} 
    \item Transactions are sent to nodes using encoded messages on the peer-to-peer network. Before geth starts processing the transaction, we save its timestamp, hash, and source node in a local database.  
    \item Once each transaction is processed, we store additional information such as the sender, receiver, and data if it was a valid transaction. This information is saved to a local database.
\end{itemize}

Then, we implemented a parser in JavaScript using the web3 library that runs
continuously in parallel to the geth node. This parser uses a multi-threaded
architecture to fetch all the metadata necessary for processing transactions:
\begin{itemize} 
\item One thread retrieves the list of memory pool transactions
from the database.  In order to avoid memory leaks, we only keep a list of
transactions within a day of the block the program is currently processing.
\item Another thread loads the list of new blocks and adds them to the work
queue to be processed. 
\item Finally, the last auxiliary thread loads the
contract Application Binary Interfaces (ABIs) for the top open-source
smart-contracts fetched from the Etherscan website \cite{etherscan}. ABIs are
the links that convert the binary inputs and outputs of smart contracts to human
readable format---this allows us to parse execution logs and understand which
events were emitted.
\end{itemize}

For each new block, the main thread performs the following steps: 

\begin{enumerate}
\item If the block has been confirmed less than 30 times, it waits to make sure
consensus has been reached.  
\item The program loads the list of transactions
and, for each transaction, runs it on a modified EVM to
retrieve all function calls, events, and internal transfers.  
\item The program
checks every transaction against the memory pool and concludes whether the
transaction is public or not. 
\item Then, for each transaction, the program uses
stored ABIs to decode the transaction inputs.  If the ABI cannot be found
locally, it is loaded from Etherscan; if absent from Etherscan, it is loaded from 4byte
\cite{4byte}.  

If the program is still unsuccessful
at finding the right ABI, it skips the current transaction and moves
on to the next one.  
\item After decoding the function, it decodes execution
logs using the contract ABI to find Transfer, Swap, Deposit, and Withdrawal
events.  These events are used to build an ordered list of transfers, including
internal ETH transfers, and ERC-20 coin or ERC-721 token exchanges.  It uses the
output from the instrumented EVM to maintain the original transfer
ordering and uses Swap events to tag pairs of transfers corresponding to a
currency exchange.  
\item If necessary, the program reorders transfers of a Swap
event to make sure the intended order is preserved.  This ensures we do not
miss cycles in the transfer graph because of out-of-order swaps.  
\item Finally,
it computes burnt fees and tips, adds these as transfers in the graph, and saves
the block in the dataset.
\end{enumerate}

\subsection{MEV Identification}

We use a separate algorithm, running offline, to identify MEV extractions in
blocks.  To our knowledge, we are the first to use a graph algorithm capable of
finding multiple types of MEV strategies in executed transactions.  Previous
works have used graph methods to find optimal arbitrage cycles in DEX rates
and identified specific types of MEV in existing blocks by looking at
account balances, but none that we are aware of implement a strategy that finds
both single and multi-transaction MEV exploits.

To do this, we came to the realization that many forms of MEV can be described
in a unified manner: MEV is a way of generating a profit from using market
inefficiencies, either created by non-synchronized exchange rates or instability caused by a large transaction.  In all cases, it is
characterized by a strategy in which, using a sequence of swaps, a bot exploits
a temporal or spatial difference in exchange rates. This can be done over a
single transaction, or over multiple. In the single transaction case: if it exploits
exchanges among different coins, it is arbitrage; if it exploits exchange rate differences
for the same coin over different DEXs, it is backrunning; finally, if it exchanges a set
of coins in a first transaction, and does the reverse exchange in another, it is
sandwiching. However, one can imagine there are more complex strategies over multiple
transactions.

Thus, we start by building the ordered graph of transitions across a block where
nodes are wallets and contracts and edges are transfers.  Doing our analysis
over a full block allows us to find complex strategies using multiple
transactions or over long chains.  


In this graph, an exploited MEV opportunity is characterized by a cycle
satisfying a set of constraints: 

\begin{enumerate} 
    \item The cycle follows the
temporal ordering of the transactions, e.g., except for the first and last edge,
all other edges are in the same order they occurred in the transaction. 
    \item Adjacent edges are either in the same currency or they are part of a currency exchange. 
    \item The cycle goes through at least two different currencies. 
    \item If one transfer of a currency exchange is in the cycle, the other must be as well. 
    \item The first and last edge must be in the same currency.
\end{enumerate}

The second constraint may seem arbitrary; however, it ensures we do not confound
MEV exploits with other forms of activity.  Sending one currency and receiving
another can happen for many reasons. Only the presence of a Swap event
guarantees \emph{intent} to exchange currencies.  This is central to MEV: the
parties with whom the bot exchanges are simply exchange platforms and are not
actively participating in extracting MEV.

If we are missing or cannot decode logs, or if transfers are out of order for an unknown
reason, constraints (1), (2), or (4) above may be violated for an MEV extraction.
However, omitting these conditions leads to many false positives. With
the current set of constraints, we did not find any false positives, despite
finding many instances of MEV.

Formally, we can view this method as the following.  Consider a set of
currencies $\mathcal{C}$ and a graph $\mathcal{G} = \left( V, E \right)$, where
the vertices $V = \left\{ n_1, \cdots, n_k \right\}$ are the Ethereum wallets
and contracts, and the edges $E = \left\{ t_1, \cdots, t_l \right\}$ are the
ordered internal transactions and coin or token transfers.  Each edge is a
triplet $t_i = (n_a, n_b, c)$ where $n_a, n_b \in V$ and $c \in \mathcal{C}$ is
a currency.  Finally, we introduce the Swap equivalence relation of edges
$\swapeq$ for which $t_i \swapeq t_j$ if and only if $t_i$ and $t_j$ are part of
the same swap exchange.  Then, an MEV cycle is defined as a cycle $L = (t_{i_1},
\cdots, t_{i_m})$ of edges satisfying:

\begin{equation*} 
    \begin{cases} 
        \forall j < k, \,i_j < i_k \\ 
        \forall j < m, \, t_{i_{j+1}}.c = t_{i_{j}}.c 
            \text{ or } t_{i_{j+1}} \swapeq t_{i_{j}} \\ 
        \forall t \in L, \, \exists t' \neq t \in L \text{ s.t. } t 
            \swapeq t' \\ 
        \left| \left\{t.c \, | \, t \in L \right\}\right| > 1. \\ 
        t_{i_1}.c = t_{i_m}.c 
    \end{cases}
\end{equation*}

Finding all loops in a graph is a problem in NP---fortunately, our algorithm
works well experimentally given the specific conditions on MEV which removes many
cycles. It took 30 minutes to process our data on a modern laptop computer.

Our process is the following.  We start by simplifying the transaction graph by
coalescing adjacent edges together with the same value and same currency.  Then,
for each new edge, we update the list of paths entering the destination node,
only keeping paths that satisfy conditions (1) and (2).  We then examine these
paths before processing new edges---if we find a cycle, we consume it, update
the graph, and continue.

The consumption process takes a cycle and finds the maximal flow that can be
supported by the cycle considering each transfer value is the capacity of the
edge. This allows us to support sub-optimal MEV strategies with leftover
capacity.

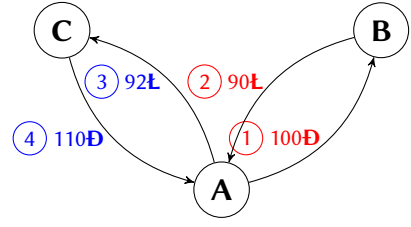
\begin{figure}
\begin{tikzpicture}[->,>=stealth',auto,node distance=3cm, thin,main
node/.style={circle,draw,font=\sffamily\Large\bfseries}] \node[main node] (A)
{A};  \node[main node] (B) [above right of=A] {B};  \node[main node] (C) [above
left of=A] {C};

\path[every node/.style={font=\sffamily\small,color=red}] (A) edge[bend right]
node [left] {\circled{1} 100\textbf{\DH}~} (B) (B) edge[bend right] node [left]
{\circled{2} 90\textbf{\L}~} (A); \path[every
node/.style={font=\sffamily\small,color=blue}] (A) edge[bend right] node [left]
{\circled{3} 92\textbf{\L}~} (C) (C) edge[bend right] node [left] {\circled{4}
110\textbf{\DH}~} (A); \end{tikzpicture} \caption{ MEV cycle example with
leftover capacity.  Order is indicated by the first index on each edge.  Edges
corresponding to the same exchange have the same color. }
\label{fig:graph-example}
\end{figure}

Consider the example given in Figure~\ref{fig:graph-example}.  This example
resembles many real sandwiching attacks.  There are three nodes $A$, $B$ and $C$, and
two currencies, \textbf{\DH}~and \textbf{\L}. $A$ exchanges 100 of \textbf{\DH}~
for 90 of \textbf{\L}~ with $B$, then exchanges 92 of \textbf{\L}~ for 110 of
\textbf{\DH}~ with $C$.

\begin{figure}
\begin{tikzpicture}[->,>=stealth',auto,node distance=3cm, thin,main
node/.style={circle,draw,font=\sffamily\Large\bfseries}] \node[main node] (A)
{A};  \node[main node] (B) [above right of=A] {B};  \node[main node] (C) [above
left of=A] {C};

\path[every node/.style={font=\sffamily\small,color=blue}] (A) edge[bend right]
node [left] {\circled{3} 2\textbf{\L}~} (C) (C) edge[bend right] node [left]
{\circled{4} 2.4\textbf{\DH}~} (A);

\path[every node/.style={font=\sffamily\small,color=black}] (A) edge [loop
right] node {7.6 \textbf{\DH}~} (A); \end{tikzpicture} \caption{ Graph from
Figure~\ref{fig:graph-example} after consumption.  The MEV cycle was removed and
replaced with a MEV gain of 7.6 \textbf{\DH}~ for A. The second exchange had
leftover capacity---the residual exchange remains in the graph. }
\label{fig:graph-example-2}
\end{figure}
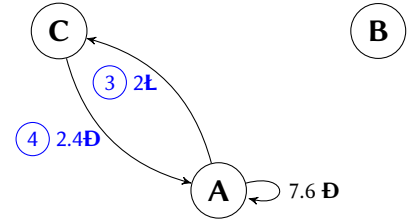

The first exchange rate is of 0.9~\textbf{\L}~ for \textbf{\DH}, while the
second is 0.836~\textbf{\L}~for~\textbf{\DH}.  It is clear that the strategy
of this loop is profitable.  However, if we only look at the balances, $A$ ends up
with 10~\textbf{\DH}, and -2~\textbf{\L}.  At the current exchange rate
between \textbf{\DH}~ and \textbf{\L}, this is still a profitable strategy; 
however, this could change if the rate evolves.  We categorize such an exchange to be
\emph{sub-optimal}: the final balance vector is not strictly superior to the initial
one.

\begin{figure*}[ht] \centering
\includegraphics[width=\linewidth]{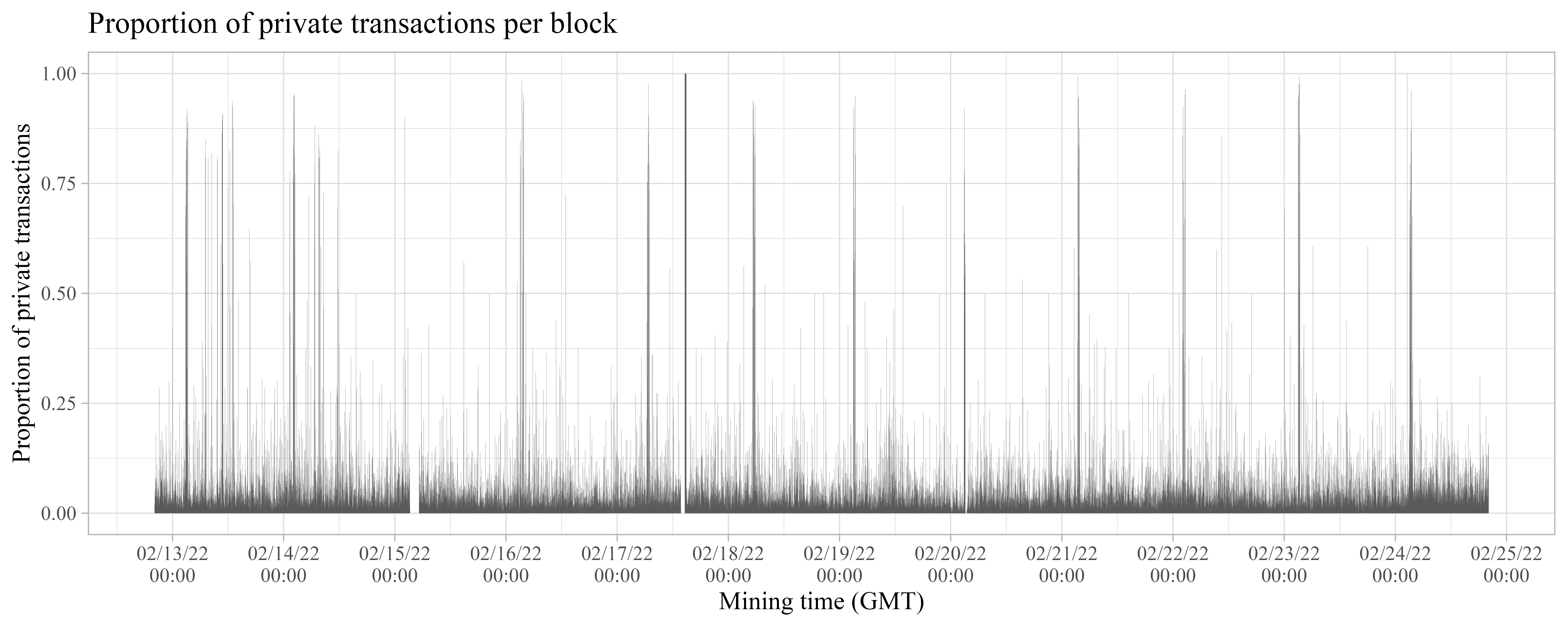}
\caption{Percentage of private transactions per block, as a function of the
mining time.\\ Note that the gaps in the figure are due to missing information,
because of errors in our data collection pipeline.} \label{fig:private-per-time}
\end{figure*}

Prior work often discarded these cases because of such ambiguity.  However, at
the instant of the MEV execution, $A$ still made a profit, although lower
than 10~\textbf{\DH}.  In order to find the true profit, we find the maximal flow
that can go through the cycle---in this case, we have to reduce the flow for the
second exchange.  We then memorize how much profit was made from that cycle, and
``consume'' it by canceling out its edges.  For instance, in the example we just
gave, after consumption, the graph will be that of
Figure~\ref{fig:graph-example-2}.  After computing the maximum flow, we find
that $A$ made a profit of 7.6 \textbf{\DH}.  The MEV cycle is thus removed from
the graph, and the second exchange is left with its residual capacity.

In reality, at the end of this cycle, $A$ is left with 10~\textbf{\DH}~ and -2
\textbf{\L}.  The profit we compute with our consumption process is equivalent
to removing the 2 \textbf{\L}~ from the 10 \textbf{\DH} after conversion at the
current exchange rate.  However, the arbitrage opportunity relies on the very
fact the exchange rate differs from one platform to another, and exchange rates
vary continuously on DEXs, so our computation is only an approximation of the
profit made by the bot. This method also assumes the exchange rate is the same
regardless of the amount transferred, which is not true in many DEXs, making
this more approximate for large transfers. Nevertheless, in practice, we found
the difference between exchange rates was small enough to have a negligible
effect on the profit and most MEV cycles we observed were optimal; hence, we did
not require such approximations.

For every MEV cycle, the algorithm records how much was gained or lost by each
of its participants, in each currency.  Then, at the end of the iteration, it
uses the conversion rates learned during the process to convert all tokens back
to ETH and coalesces cycles that are part of a larger common structure.
As stated previously, using learned conversion rates is an approximation for the
total profit, however most MEV gains are directly in ETH, and the difference
between conversion rates was small enough to have negligible effect on the
result.

To find the miner's compensation, the algorithm then looks at burnt fees, gas tips, and hidden transfers to the miner for each MEV cycle.

Using this approach, we find MEV extractions, categorize them, and keep track of
how much was earned by the user who executed the corresponding transactions as well as how much was sent to the
miner in payment.  We processed all of our data and found a total of 29,571 MEV
extractions, which we analyze in the following section.

\section{Results}
\label{sec:results}

We now review our results. We start by discussing private transactions, before 
analyzing profit redistribution from MEV and the impact it has on the 
stability of the blockchain. 


\subsection{Private Transaction Usage}

Private transactions are rare events in the Ethereum blockchain---only
\privateoverall~of all observed transactions in blocks did not appear in our
memory pool. Even rarer are Flashbots transactions, which only appear in
\flashbotoverall~of all observed transactions. However, private transactions are
not distributed uniformly across miners or blocks. In fact, when plotting the
percentage of private transactions per mined block as a function of the
validation time, in Figure~\ref{fig:private-per-time}, we see spikes of private
transactions every day around 3AM GMT.  After tracing the blocks involved, we
found these spikes are due to mining profit re-distributions from the
\texttt{F2Pool} mining pool. In order to avoid paying fees to other miners,
some pools include minimal gas ETH transfers to pool members in their own blocks
without submitting these transactions to the memory pool.  In general, spikes in
this graph correspond to blocks with a majority of private transactions: this is
characteristic of profit re-distributions.

\begin{figure*}[h] 
    \centering
    \includegraphics[width=\linewidth]{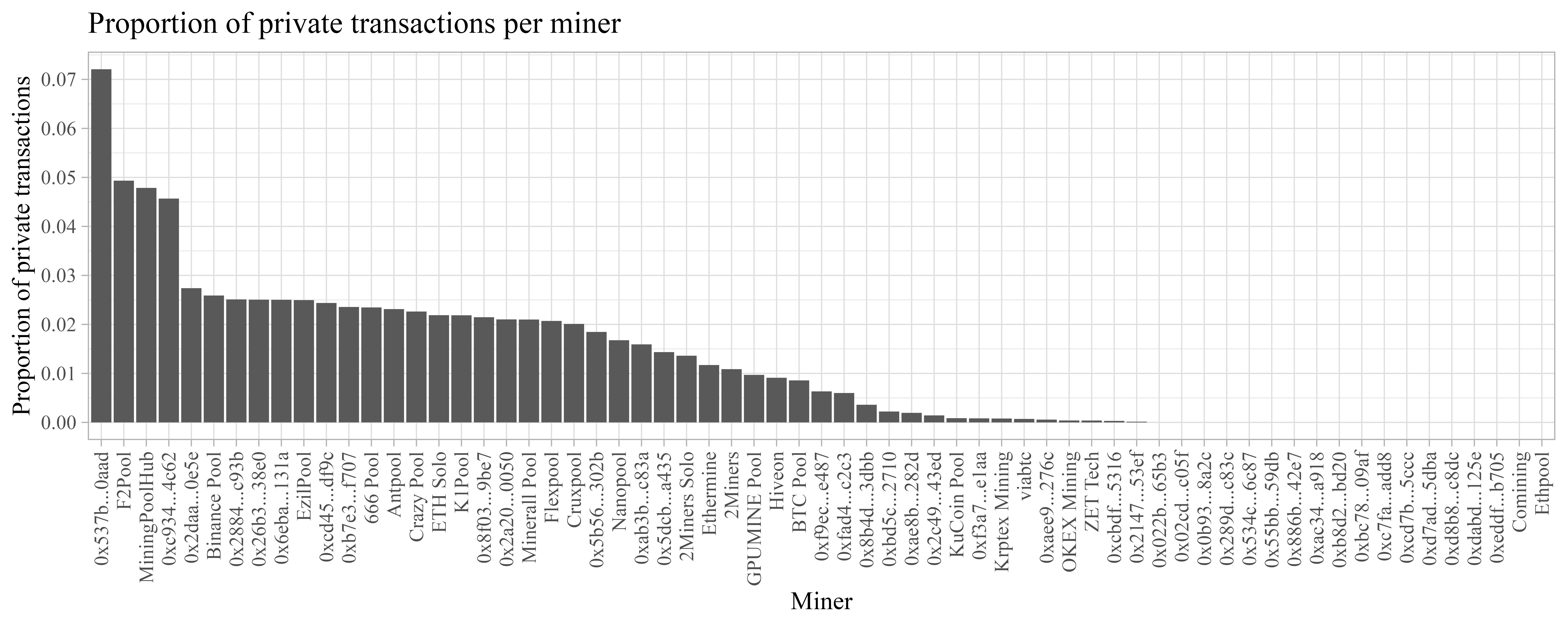}
    \caption{Percentage of private transactions for blocks mined by each miner.}
    \label{fig:private-per-miner} 
\end{figure*}

Private transactions are not used by all miners equally.
Figure~\ref{fig:private-per-miner} shows the proportion of private transactions
in blocks mined by specific miners.  The first notable finding is that about
40\% of miners in our dataset include less than 0.1\% of private transactions in
their blocks. Considering our local memory pool might miss some transactions, it
is likely these miners in fact do not use any private transactions, and these
extremely low values are noise from the measurement process.


However, this figure also shows some miners rely heavily on private
transactions. Four miners included over 4\% of private transactions---double
the average of \privateoverall.  These are, in order of prevalence,
\texttt{0x537b}\dots\texttt{0aad}, with 7.2\% of private transactions, miner
\texttt{F2Pool}, with 4.9\% of private transactions, \texttt{MiningPoolHub},
with 4.8\% of private transactions, and \texttt{0xc934}\dots\texttt{4c62} with
4.6\% of private transactions.

\begin{figure}[h]
    \centering
    \includegraphics[width=\linewidth]{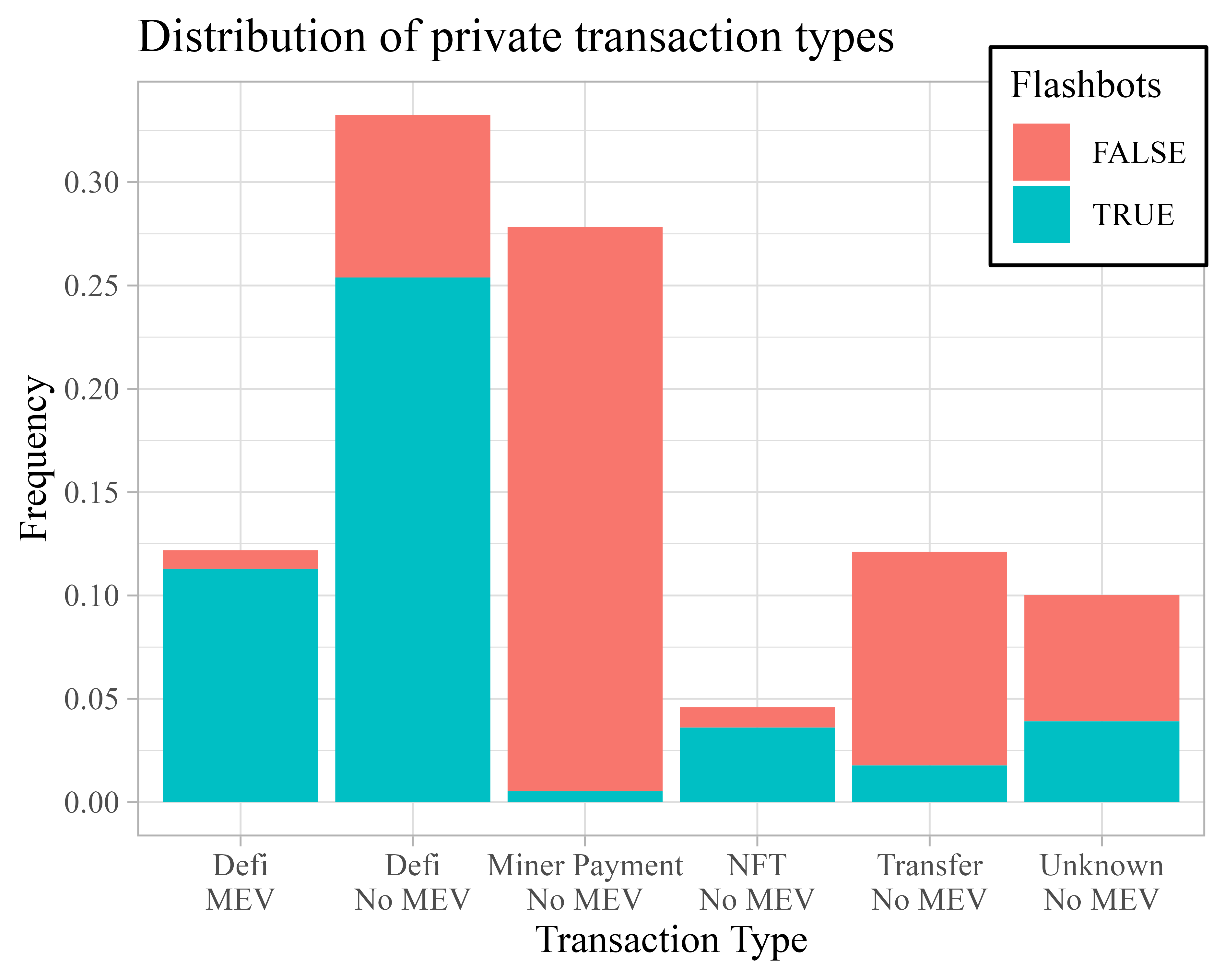}
    \caption{Distribution of private transactions per category.}
    \label{fig:type-per-private}
\end{figure}

We used our data to understand what private transactions are used for in order
to better grasp these differences across miners.  We remark that profit
re-distributions can explain some of this disparity---participation in the
Flashbots network is one possible explanation.  First, we plot the
distribution of transaction categories for private transactions in
Figure~\ref{fig:type-per-private}.  The categories are the following:
\begin{description} \item[Defi.] Transactions that exchange ERC-20 tokens.
\item[NFT.] Transactions that exchange ERC-721 tokens. \item[Miner Payment.]
Miner profit re-distributions. \item[Transfer.] ETH transfers without smart
contract input data. \item[Unknown.] Interactions with smart contracts that do
not involve ERC-20 or ERC-721 tokens.
\end{description}

Furthermore, we separate each category into two subcategories: those for
which we identified MEV extraction, and those without.  Naturally, all
identified MEV extraction occurs in DeFi, since our underlying definition of MEV
stems from the mechanisms of token exchanges.  Finally, the figure indicates
which proportion of each category was seen in a Flashbots bundle.

This figure illustrates two main findings. First, private transactions are most used 
for DeFi applications and miner re-distributions. Second, Flashbots only represents 
\percentageprivateflashbots~of private transactions. However, Flashbots usage 
is not uniform across all categories. It is heavily used for DeFi---both
to frontrun and to \emph{avoid} frontrunning---and NFT, while rare for transfers and
miners payments. 

\begin{figure}[h]
    \centering
    \includegraphics[width=\linewidth]{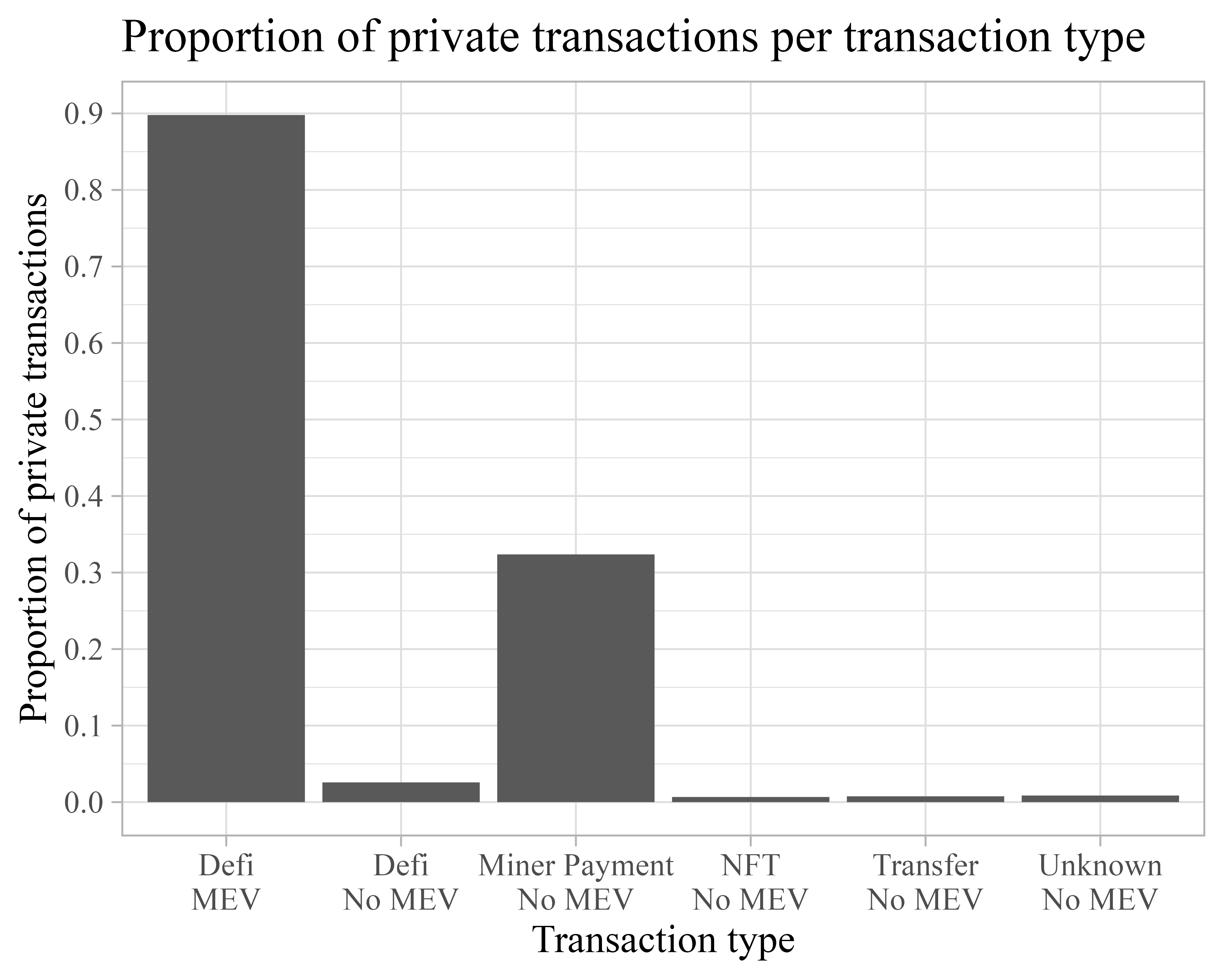}
    \caption{Prevalence of private transactions per category.}
    \label{fig:private-per-type}
\end{figure}

Next, we look at the prevalence of private transactions per category in
Figure~\ref{fig:private-per-type}.  Private transactions seem to be most used in
miner profit re-distributions (32.4\%) and MEV extractions (89.8\%), while only
rarely used for NFTs (0.7\%) or regular transfers (0.8\%).  This appears to be
one of the major consequences of the Flashbots system \cite{flashbots}, which
allowed MEV to be extracted over private channels, thereby sidestepping public
priority gas auctions.

Thus, we find that private transactions are not uniformly used and are the
preferred mechanism for profit redistribution and MEV exploitation.  We believe
the increased usage of private transactions enabled by MEV relays weakens the
transparency and fairness of the blockchain as a whole.

Fairness is impacted because of the information asymmetry resulting from private
transactions.  Users that directly communicate with miners are given the
advantage of knowing other users' transactions (by reading the memory pool)
without revealing their own. Private transactions have been used for ethical reasons, 
such as retrieving at-risk value in poorly designed smart contracts as shown in 
\cite{DarkForest2}, but it has also often been used for theft. For example, the ChainPort security
breach in July 2021 resulted in the theft of 1.3 million USD stolen through
private transactions \cite{chainport} to make sure the fraudulent transaction
could not be seen until included in a block.

This leads us to our second point: private transactions reduce blockchain transparency.  
Without them, miners are incentivized to order
transactions by decreasing gas price, giving users a good indication on how much
gas to provide in order to be included in a block at a certain position.
However, private agreements between miners and users often lead to out-of-order
transactions since compensation can be given out-of-band.  This implies regular
users have less visibility on the outcome of their transaction as it becomes
increasingly hard to predict the order in which transactions will be processed.

However, these are not the only risks caused by private transactions and MEV
relays: they incentivize MEV extraction by eliminating PGAs, giving a false sense of fair profit redistribution.
Our data suggests that miners are making large profits by charging hefty fees to
MEV extraction bots and by running their own MEV extractions.
These profits are large enough to warrant forking attacks by miners with high
hash rate in order to steal MEV profits in other blocks.

\subsection{MEV is a Miner's Game}

Despite the rampant exploitation of MEV in today's DeFi community, not much is
known about \emph{how} these profits are redistributed.  MEV relay protocols
claim to enable fairer redistribution of MEV opportunities, however we claim
miners are making large profits for themselves, and even run their own MEV
extractions at an alarming rate.  Over the course of \totalduration, at least 53
blocks had miner rewards from MEV extraction higher than the block reward, and
two blocks had a miner reward over four times more than the block reward.  This
means that the blockchain is currently at risk of forking attacks from its highest hash rate miners.

\begin{figure*}[h] 
    \centering
    \includegraphics[width=\linewidth]{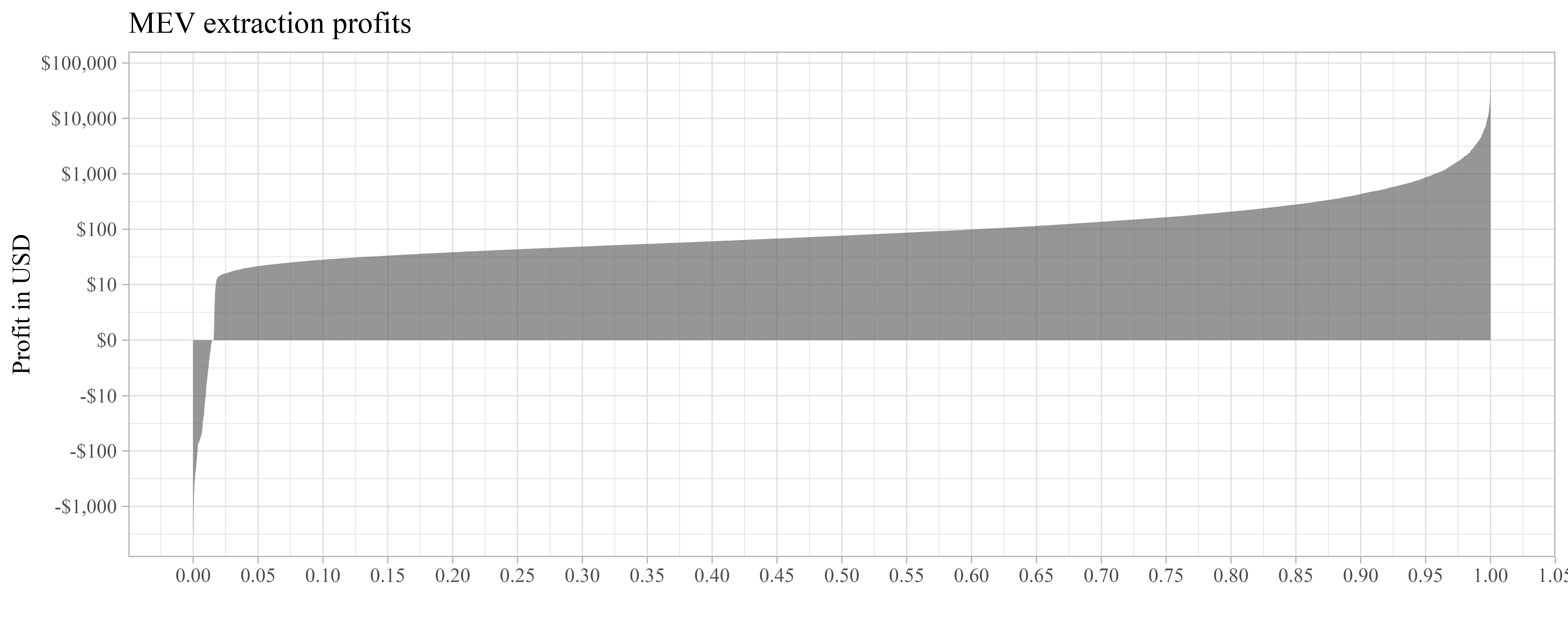} 
    \caption{Profits from all MEV extractions in increasing order.} 
    \label{fig:profit-curve}
\end{figure*}

We start by analyzing MEV profit distribution.  Figure~\ref{fig:profit-curve}
shows the profits from every MEV extraction in our dataset in increasing order.
Two trends stand out:

\begin{itemize}
    \item About 2\% of all MEV extractions resulted in a loss.  This highlights
    the time sensitivity of arbitrage strategies. In fact, if we only consider
    MEV extractions in our data that had negative returns after burnt fees, only
    43.9\% used Flashbots even though the global average for MEV was 86.6\%. This
    demonstrates that MEV relays simplify capture of MEV opportunities by eliminating
    gas ordering constraints.  However, these failures are not only due to
    ordering inconsistencies---they are often due to mistakes from the MEV
    extraction bot.  A notable example can be found in block 14,206,323:
    \begin{itemize} \item Transaction \#58 wanted to exchange 59.4 ETH for the
    LOOKS ERC-20 token.  \item An arbitrage bot saw the opportunity for a
    sandwiching attack and bought 34.7 ETH worth of LOOKS in transaction \#57.
    \item Unfortunately, this impacted the exchange rate for LOOKS, which made
    the first transaction fail. \item When the bot went to sell his tokens, he
    did not have the benefit of the higher exchange rate caused by the victim
    transaction. This triggered a loss of 0.43 ETH.
    \end{itemize}

    \item Most MEV extractions make profits between 10 USD and 500 USD.  Considering
    that we find on average 2,487 profitable MEV extractions per day, this
    amounts to large profits.  Keeping into account the negative MEV
    extractions, we find that MEV generates 187 ETH per day after burnt fees,
    or about 560,000 USD at the time of data collection. The largest single MEV
    extraction happened in block 14,258,520, in which a bot made an
    estimated profit, using the maximal flux technique from
    Section~\ref{sec:experimental} of 20.6 ETH (around 61,800 USD) from an
    arbitrage attack.

\end{itemize}

In total, the MEV extractions in our data generated 2,159 ETH (about
6,400,000 USD), only over \totalduration.  As a measure of comparison, this
represents 2.2\% of the total ETH supply created during that same time, and
extrapolates to almost 200,000,000 USD in profits per year.

63.2\% of those profits come from sandwiching attacks, the most notable example
being in block 14,217,123 where a bot made a profit of 16.9 ETH by exchanging
ETH with DAI. Next, 20.8\% are a result of backrunning, with the largest
extraction in block 14,260,017, where a bot took advantage of a pricing
difference between SushiSwap and Balancer caused by an earlier transaction.
Finally, arbitrage makes up the last 16.0\% of the profits, the most notable
example being given previously. \\

However, a large part of these profits is kept by the miners. It is common for
bots to include large gas fees in their transactions as a way to be placed
first in the block and as payment to the miner when using private transactions.
Sometimes, these transactions explicitly transfer part of the profits to the
miner using internal transactions. We thus consider all gas fees and miner
transfers in MEV extraction transactions to be the miner's profit.  This is most
likely an underestimation, since miners could be paid out of band by the bot,
or could control the bots and choose to keep the balance on the bot addresses.
However, we can never be sure a subsequent transfer to a miner constitutes a
payment for exploiting the MEV. This is because there is no direct link between
the MEV extraction and the transfer. We thus keep our conservative definition as
our estimate for miner profits, inferring the real value is higher.

In our data, MEV income represents \minerpercentageincome~of a miner's
total income from tips and transfers, and \minerpercentageincomeDeFi~when
restricted to income from DeFi transactions.  

\begin{figure}[h] \centering
\includegraphics[width=\linewidth]{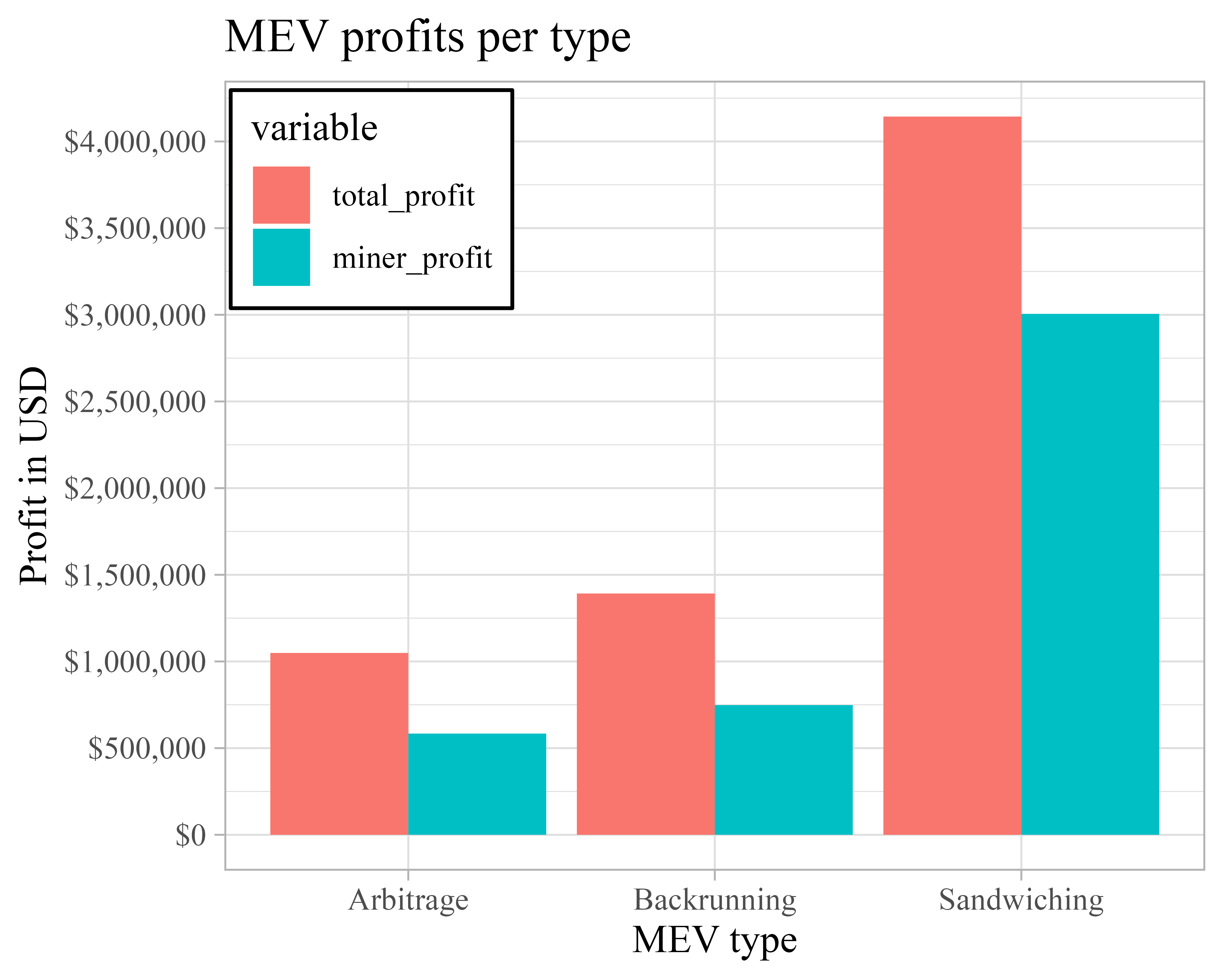} \caption{Miner
and total profits per MEV type.} \label{fig:profit-per-type}
\end{figure}

Miners made 1,456 ETH (approx. \$4,370,149) in a week, representing
\minerpercentageprofit~of all MEV profits after burnt fees, making miners the
largest MEV beneficiary.  We break down miner profits per MEV type in
Figure~\ref{fig:profit-per-type}.  Miners extract most from sandwiching by
taking 72.5\% of all profits, then from arbitrage (55.7\% of all arbitrage
profits), and finally backrunning (53.7\%).  This inequality in MEV profit
redistribution undermines Flashbots' claim to fairness; it raises the concern 
that miners are independently extracting MEV in addition to claiming large 
proportions of bot extractions.

\begin{figure}[h]
    \centering
    \includegraphics[width=\linewidth]{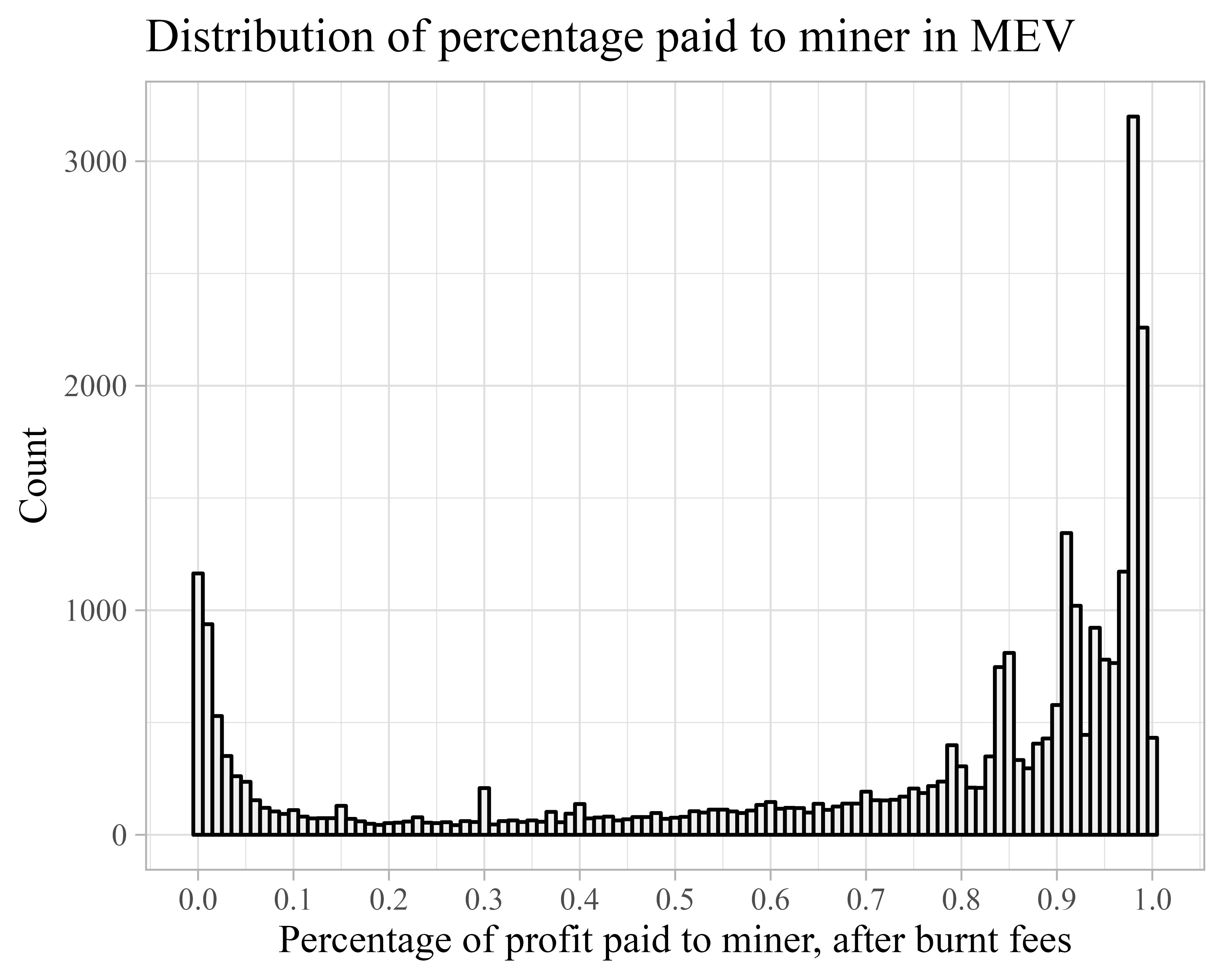}
    \caption{Distribution of profit percentage paid to miner in fees.}
    \label{fig:profit-percentage-dist}
\end{figure}

In order to investigate this possibility, we plot the distribution of miner
fees in MEV transactions as a percentage of the total profit extracted in
Figure~\ref{fig:profit-percentage-dist}. We find several noteworthy trends:

\begin{description}
    \item [No gas fee.] At the very left of the graph, a significant proportion
    of MEV extractions paid no gas fees and sent no internal transfers to the
    miner.  This situation most likely indicates one of two possibilities:
    \begin{enumerate*}[label=(\roman*)] 
    \item The bot that exploited the MEV
    colluded with the miner but paid the miner out-of-band for being included in
    the block. 
    \item The miner itself performed the arbitrage and kept all of
    the profits on the contract that performed the MEV. 
    \end{enumerate*}

    \item [Low gas.] Some transactions paid a very low percentage of their
    profits to the miner.  These were bots which most likely paid standard or
    slightly higher than standard gas prices while making profit.

    \item [High gas.] Interestingly, the highest frequency count is very close
    to 100\%, meaning many MEV extractions sent all their profit to the miner.
    We conjecture that these bots were controlled by miners because the leftover
    profit was often negligible, if any. After analyzing some transactions in
    this area of the graph, we realized the same contracts appeared often and were
    shared across different mining pools for profit: the same bot was used by
    multiple miners.

    \item [Spikes.] Finally, we investigate the four intermediate spikes. 
    They corresponded to activity from a specific contract that most
    likely sent a fixed percentage to the miner as compensation.
    \begin{description} 
        \item [30\%.] This spike was created by contract
        \texttt{0x0000}\dots\texttt{6b40} which exploited both arbitrage and
        sandwiching opportunities.  
        \item [80\%.] Caused by contract
        \texttt{0x7cf0}\dots\texttt{604f} which transferred most of its earnings to a
        set of mining pools. 
        \item [84\%.] Caused by contract
        \texttt{0xe33c}\dots\texttt{ea70} which only interacted with one address, and
        exclusively paid miners using gas fees instead of transfers. 
        \item [94\%.]
        Similar to the 80\% spike, this one was due to \\
        \texttt{0x0000}\dots\texttt{594e} which also transferred profits to a set of
        miners.
    \end{description}
\end{description}

The mode at 98\% in the graph seems to suggest many MEV extractions are in fact
directly coordinated by miners instead of third party bots.  We chose to
classify extractions with miner fees above 95\% of the total profit as
\emph{miner-driven} and other profit as \emph{bot-driven}.  
Although this is speculative, we find that miner-driven
MEV extractions were made up of 97\% of private transactions, while bot-driven
extractions were only made up of 79\% of private transactions, suggesting our categorization
is warranted.

\begin{figure*}[ht] 
    \centering
    \includegraphics[width=\linewidth]{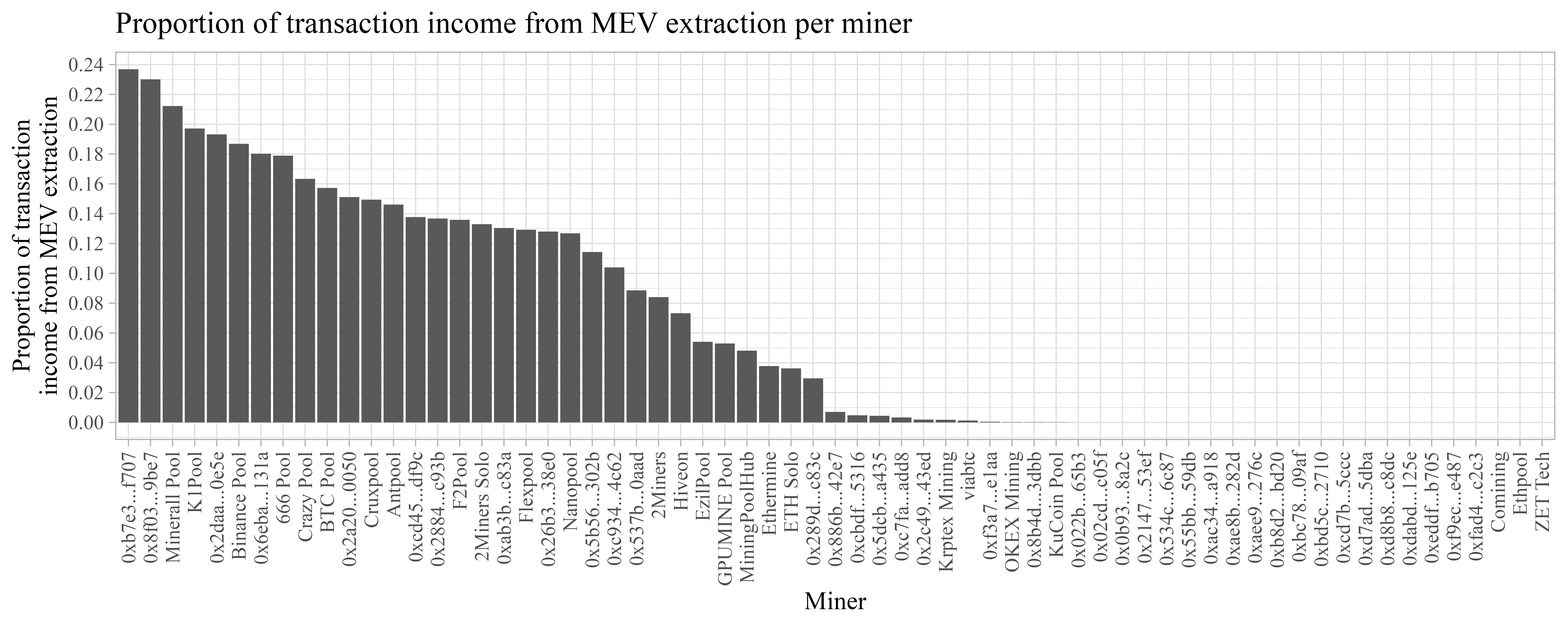}
    \caption{Proportion of miner profit from MEV per miner.}
    \label{fig:profit-per-miner} 
    
\end{figure*}

\begin{figure}[h]
    \centering
    \includegraphics[width=\linewidth]{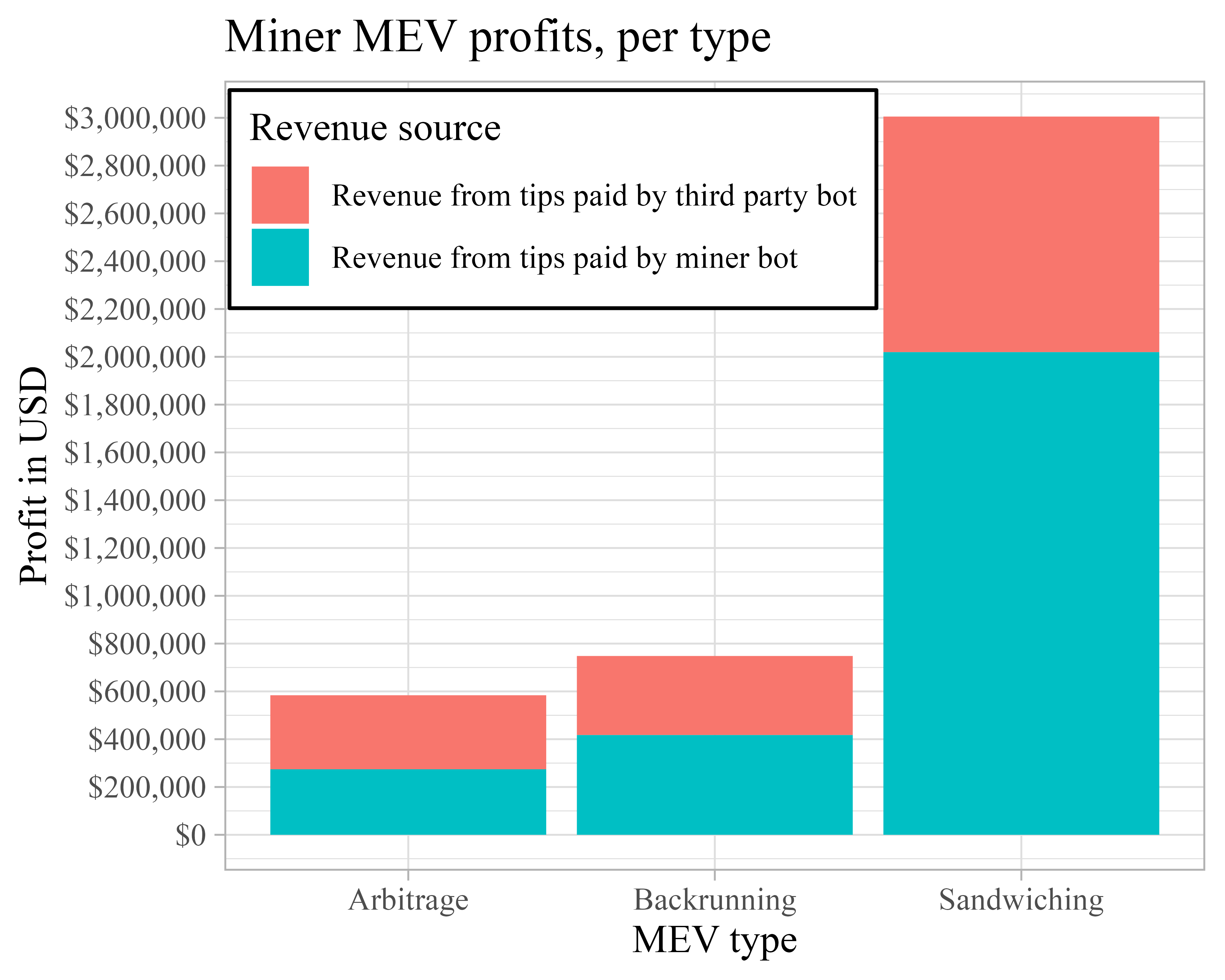}
    \caption{Miner profit broken down by MEV type and MEV extractor.}
    \label{fig:miner-profit-per-type}
\end{figure}

As an example of such miner-driven MEV extraction, consider the sandwiching attack in block 14,195,580 at transactions 0 and 2.  For the first
part of the sandwich, in transaction 0, the bot paid no gas tip to the miner.
However, in transaction 2, the bot paid 5.92 ETH to the miner in gas tip while
making a profit of 6 ETH: 98.6\% of the profits went directly to the miner.
The scenario in this example was common to all miner-driven sandwiching attacks and was the original finding that reinforced confidence in our classification of miner-driven MEV.

Now, breaking down miner profits from ``miner-driven'' and ``bot-driven''
sources in Figure~\ref{fig:miner-profit-per-type}, we find that the majority of
a miner's profit comes from bots in its control. \\

As for private transactions earlier, not all miners participate in MEV
extraction.  Figure~\ref{fig:profit-per-miner} breaks down profits from MEV
relative to total income from tips and transfers per miner.  40\% of miners do
not take part in MEV extractions, whereas some make almost a quarter of their
total transaction fee profit directly from MEV extractions.  Unsurprisingly,
miners with a high MEV income proportion often also have a high rate of private
transactions.  For example, miner \texttt{0x2daa}\dots\texttt{0c5c} is the fifth
highest both in number of private transactions and in relative profit from MEV
extraction.

The miner with the most profit from MEV is \texttt{0xb7e3}\dots\texttt{f707},
with 24\% of MEV-driven profits.  The profitability of MEV for miners is even
more blatant if we only consider profits from DeFi:
\texttt{0xd2aa}\dots\texttt{0e5e} makes 39\% of its DeFi transaction 
profits from MEV.

\begin{figure*}[h] \centering
\includegraphics[width=\linewidth]{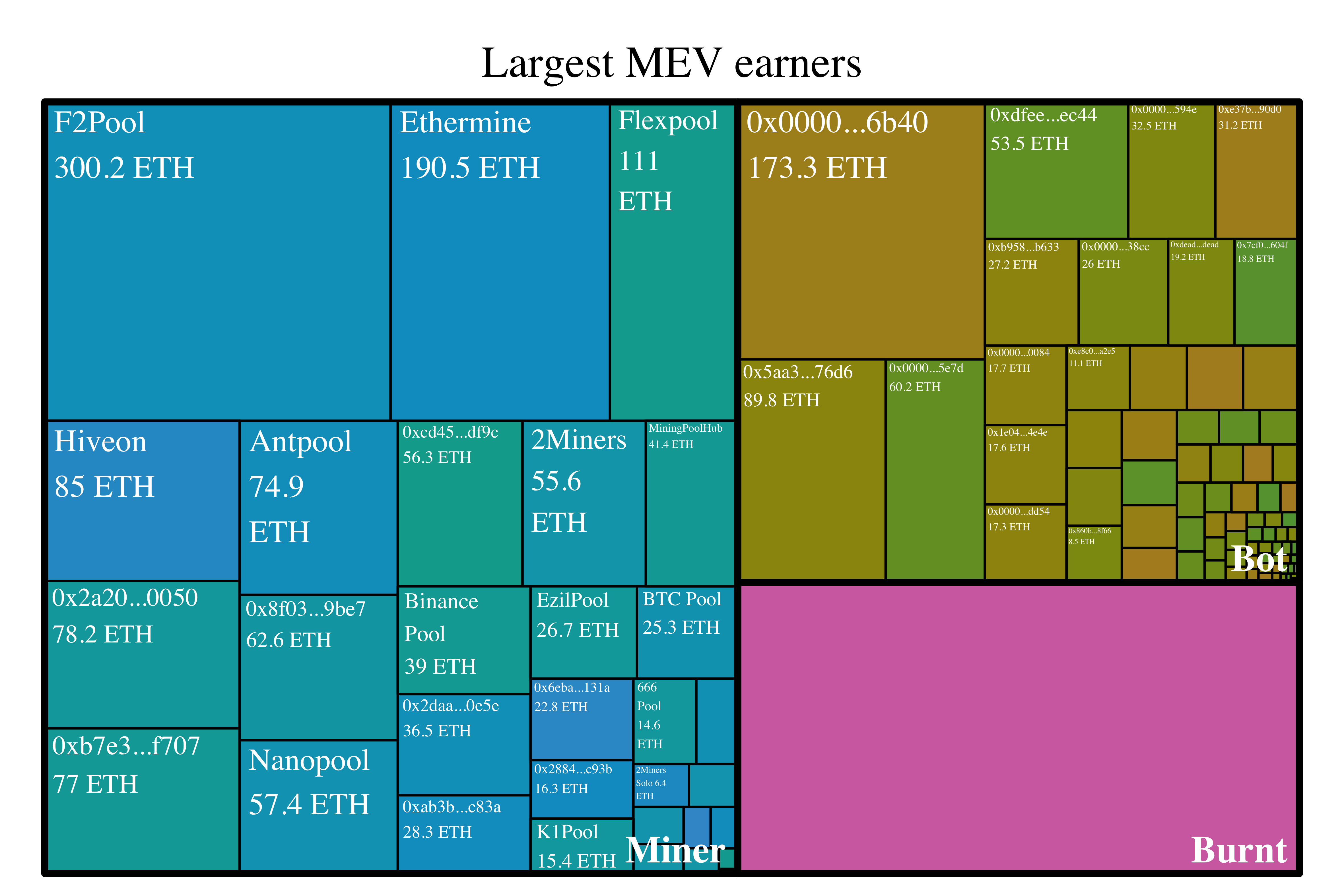} \caption{Largest
individual earnings from MEV.} \label{fig:treemap} \end{figure*}

In order to understand the scale of miner's domination over MEV, we map the
total MEV earnings per address in Figure~\ref{fig:treemap}.  The top five miners
earn more than all bots combined, and the top 10 miners make over half of all
profit.

Estimating total gains for individual bots is challenging because tracking final
profit recipients is difficult.  We slightly overestimate bot profits in this
graph: our methodology is to subtract miner and burnt fees from raw MEV profits
generated by each bot without considering other fees such as flash loan interest
payments.  This hints that the contrast between miners and bots could be even more
stark in practice.


We believe these findings invalidate Flashbots' claim to fair redistribution of
MEV.  At most, the redistribution is fair among collaborating miners.

However, miner domination of MEV is concerning for reasons other than fairness.
First, as in any distributed ledger without a central trust authority, miners are
validators---they enforce the rules set forth by the Ethereum protocol.  They
serve as the regulation authority, ensuring the properties of transparency and
fairness in Ethereum.  Thus, having up to a quarter of their transaction income
come from exploiting market inefficiencies, many times at the expense of regular
users (for sandwiching attacks, for instance), represents a major conflict of
interest that impedes the ability to serve all users equally.

Second, as emphasized in the original Flash Boys 2.0 paper
\cite{daian2020flash}, large MEV profits incentivize miners to perform
time-bandit attacks.  These attacks happen when a miner decides to mine a
previous block themselves to claim profits or change transactions. Afterwards,
they proceed to mine the next blocks until the new chain is long enough to
convince other nodes to build on it.  This can allow the miner to recuperate the
profits from MEV in an older block, but it can also lead to double-spending.
However, orchestrating such an attack is, in general, not economically viable
due to the small chance of success. In fact, \cite{zhou2021just} used a
Markov Decision Process to model this probability.  They found that with the
current stale block rate, miners with a hash rate greater than 10\% of the total
Ethereum hash rate had positive profit expectation if they produced more than four
times the block reward. 
In our data, we found two
miners with a hash rate greater than 10\%: \texttt{Ethermine} at 35\%, and \texttt{F2Pool} at 14\%.
Both these miners have a history of profiting from MEV opportunities.  \texttt{Ethermine}
made 190 ETH in a week in our data, representing 3.8\% of its profits from gas,
and \texttt{F2Pool} made 300 ETH, or 13.6\% or its profits from gas.

We found two blocks in our 12 day span, 14,217,123 and 14,241,282, for which
miner profits from MEV exceeded four times the block reward. The block reward
here is the sum of the static block reward (2 ETH) and the fees from all non-MEV
transactions. Forking these blocks would have simply required mining them
without changing their content, since the miner fees from MEV alone exceeded
four block rewards.  Furthermore, we found four blocks for which the total MEV
profit (not simply the part shared with the miner) exceeded four block rewards.
These could have been forked as well, with the additional constraint of
replacing MEV transactions with the miner's own transactions.  We did not find
traces of fork attacks in our data; however, it is only a matter of time before
they start happening, since there are multiple opportunities for forking every
week. This represents an existential threat to the stability of Ethereum, and
to the trust users hold in its potential.

\section{Discussion}

\label{sec:discussion}


\paragraph{Limitations of existing countermeasures.} 
Other than Flashbots Auctions which, as we have seen, does not mitigate most MEV risks,
other countermeasures have been proposed as covered in
Section~\ref{sec:related}.  
However, to our knowledge, none today are satisfactory for mitigating the stability risks incurred by MEV.  
A line of work \cite{heimbach2022eliminating,mcmenamin2022fairtradex,ciampi2021fairmm,baum2021p2dex, zhou2021a2mm} focus on better DEX constructions, and in particular, AMM design. 
These papers, however, fail to capture the full extent of MEV.
In most cases, inter-DEX arbitrage is still possible, and only a few address the price slippage issues that make sandwiching attacks possible.


Research into ensuring order-fairness in transactions seems more effective since it
prevents reordering and mitigates order insertion, both of which are root causes
for MEV.  However, strict arrival order fairness is impossible, as shown by
\cite{kelkar2020order}.  Instead, papers rely on weaker versions that do not
prevent all MEV attacks.  \cite{kelkar2020order, kelkar2021themis} attain batch
order-fairness, which states that if a majority of honest participants received
$t_a$ before $t_b$, then $t_a$ must be in an older block or the same block as
$t_b$.  This, unfortunately, does not prevent reordering within a block, implying
MEV extraction is still possible.  \cite{kursawe2020wendy} achieves
order-linearizability, meaning that if \emph{all} honest participants receive
$t_a$ before any receives $t_b$, then $t_a$ must be ordered before $t_b$.
Unfortunately, due to propagation time in the network, it is still possible for
an adversary to send their transaction before all nodes have received $t_a$, and
thus, can still be placed higher than $t_a$.

Finally, Helix \cite{yakira2021helix} seems to offer the best guarantees of any
system. However, as mentioned earlier, it requires a complete change of the
consensus protocol and does not prevent validators from adding their own
transaction at arbitrary positions in the block.

\paragraph{Randomizing transaction order.} 
We believe a
system based on random transaction ordering might mitigate most MEV issues.
Instead of committing a block directly, validators could first commit to a set
of transactions with any arbitrary order.  Then, using the hash of the
commitment, as well as the hash of the previous block, they could seed a random
permutation using a public inefficient function.  Using the commitment hash as
part of the random seed helps ensure the permutation was not pre-generated, and
using an inefficient function helps avoid enumerating many possible commitments
before finding the desired outcome.  Then, validators could use this permutation
to order transactions in the block.

This sketch has some flaws: adversaries with enough computational power might
still try and generate many commitments until finding one that produces a
suitable permutation.  Furthermore, bots could submit many different copies of
the same transaction to maximize their chances of being processed high in the
block.  Lastly, this protocol might drive up gas prices for small transactions
since miners will select the set of transactions with the highest payoff.
However, absent these issues, it guarantees a random transaction ordering, which
makes MEV prohibitively difficult to extract for both bots and miners.
We leave development of a practical and secure solution based on random
transaction ordering to future work.

\paragraph{On Flashbots.} The Flashbots ecosystem strives
to address the MEV issue by bringing the MEV extraction process to public
view. The idea is that ``democratized'' MEV extraction will help reduce
information asymmetries on the Ethereum blockchain. While Flashbots has its
benefits (e.g., it helps to prevent network congestion), we find that a core
assumption is violated in practice: miners are exploiting their ability to
inspect transactions to conduct individual MEV extraction via standard
techniques. By additionally participating in the Flashbots relay network,
miners gain an additional revenue source via payments. This behavior
contradicts the fairness property that Flashbots is intended to provide. We
note that Flashbots is an active project and the system is currently not
designed to provide all target properties. For instance, the \emph{current}
version of Flashbots v0.4 \cite{flashbots} does not provide full transaction
privacy, permissionlessness, or even
finality---these are the properties that directly address the MEV issue, i.e.,
the lack of these properties is what miners leverage to conduct MEV
extraction, thereby increasing information asymmetry in Ethereum. 

On a more fundamental level, Flashbots is not designed to provide \emph{order fairness}.
A number of avenues have been explored to add privacy to Flashbots \cite{miller2021mev}, potentially providing order fairness guarantees.
We summarize the proposals below.

First, searchers could withhold sending transaction details to miners, only sending the block header.
Miners would then validate the block header, and once verified, transactions could be revealed and added to the block.
This has the drawback that adversarial searchers could pass invalid blocks to miners.
Slashing has been proposed as a way to punish such misbehavior, though this would come at the cost of usability.
Second, secure enclaves such as Intel SGX would directly add privacy with little additional overhead.
The main issue with secure enclaves is the \emph{reliance} on enclave technology: such technology is not supported by rigorous security proofs or audits and new attacks are commonplace.
The main cryptography-based solutions use timelock puzzles or threshold encryption.
Timelock puzzles allow for privacy that depends on the delay introduced from the puzzle itself: by enforcing that decryption takes a fixed amount of time, the puzzle allows miners to mine on encrypted transactions that are only ``revealed'' after successful block mining.
This comes at the cost of network delay---the network has to wait for the timelock puzzle to be solved before transaction contents are visible, which again degrades network usability.
Finally, threshold encryption can enable block decryption if sufficiently many miners, each holding a unique key, attempt decryption.
If less miners than the threshold attempt decryption, it will fail.
This relies on an \emph{honest majority} of miners in the committee---if a majority of miners collude and share secret material, they can break encryption.
Such a reliance makes it difficult to use threshold encryption in a permissionless setting.
We refer the reader to \cite{miller2021mev} for a deeper summary of the trade-offs between these proposals.

\paragraph{On the Switch to Proof-of-Stake}

Ethereum is scheduled to merge with the beacon chain proof-of-stake system 
before the end of 2022. After ``The Merge,'' the network will no longer be secured by miners.
Instead, security will stem from validators that stake security deposits to
compute vote-based consensus. Validators need to stake at least 32 ETH and are
chosen at random for block validation in proportion to their stake.  This
approach leverages economic incentives that reward good validator behavior
(following the protocol) and punish bad validator behavior (deviating from the
protocol).

In Ethereum 2.0, MEV extraction is conducted by block proposers as opposed to
miners in proof-of-work Ethereum. The core reason for this is that transaction
ordering is unchanged for the proof-of-work component of Ethereum 2.0. Hence,
MEV opportunites may continue to exist in their current form, however the
entities capable of MEV extraction have changed. Specifically, they have shifted
from miners to validators. For this reason, existing MEV extraction mitigation
techniques may be portable to Ethereum 2.0---we discuss Flashbots in this light
below.


Part of the reason MEV exploitation is rampant today is due to the
centralization of mining power by a handful of miners: the top three mining
pools hold over 50\% of the hash rate. Centralization helps coordinate MEV
extraction through private relays such as Flashbots. In this regard,  proof-of-stake still concentrates validation powers in the hands of a few. Although it
has a longer tail of validators, as of today, the top two staking pools hold
over 50\% of all stake, concentrating power in the hands of even fewer
addresses.

\section{Conclusion}
\label{sec:conclusion}

We carried out careful investigation of private transaction use and MEV
exploitation on the Ethereum blockchain, using a bespoke generalized MEV
detection tool based on cycle detection in transfer graphs. We found that MEV
exploitation today is done over private transactions in 91.5\% of the cases, and
using Flashbots in 88.1\% of cases. However, despite Flashbots' hopes of
mitigating negative MEV externalities and fair access to opportunities,
\minerpercentageprofit~of MEV profits are made by miners, with the five largest
MEV-exploiting miners making more than all bots combined. Overall, MEV accounts for \minerpercentageincome~of miner profits from transaction fees and
\minerpercentageincomeDeFi~of profits from DeFi trades. However, not all miners
participate in MEV extraction equally. 40\% of them do not include any private
transactions in their blocks and make negligible profit from MEV exploitation
on their blocks, whereas one miner made 39\% of their DeFi transaction income
from MEV.

This presents serious risks to the stability of the blockchain. On average, one
block every week contains enough MEV profit \emph{in transfer fees alone}
to make a time-bandit attack economically viable for miners with a hash rate
above 10\%. The top two miners, \texttt{Ethermine} and \texttt{F2Pool}, both have a hash rate
above 10\% and have a history of exploiting MEV. In particular, \texttt{F2Pool} is the
largest single entity in terms of MEV profits, making 300 ETH of profit in
\totalduration, which accounts for 13.6\% of its total transaction profit.

Today's MEV mitigation solutions are not sufficient to address the issue. We
believe a random transaction ordering system could potentially make MEV
extraction prohibitively hard, however this remains to be investigated. Without
a proper solution, Ethereum remains an impractical blockchain for real-world
use. Regular Ethereum users will continue to pay high gas fees because of bots,
lose money to MEV schemes when trading tokens, and most importantly, face the
risk of time-bandit attacks.


\bibliographystyle{acm}
\bibliography{references}


\end{document}